\documentclass[preprint,aps,prab,amsmath,amssymb,superscriptaddress]{revtex4-2}
\usepackage[utf8]{inputenc}
\usepackage{graphicx}
\usepackage{hyperref}
\hypersetup{
	hidelinks,
	colorlinks=False,
}
\usepackage{multirow}

\begin{document}

\title{Longitudinal phase space diagnostics with a nonmovable corrugated passive wakefield streaker}
\author{Philipp Dijkstal}
\altaffiliation{Present address: Paul Scherrer Institut, Forschungsstrasse 111, 5232 Villigen PSI, Switzerland}
\author{Weilun Qin}
\author{Sergey Tomin}
\email{sergey.tomin@desy.de}
\affiliation{Deutsches Elektronen-Synchrotron DESY, Notkestr.\ 85, 22607 Hamburg, Germany}
\date{\today}

\begin{abstract}
	Time-resolved diagnostics at Free-Electron Laser (FEL) facilities, in particular electron beam longitudinal phase space (LPS) and FEL power profile measurements, provide information highly valuable for users, machine development studies, and beam setup.
	We investigate the slice energy resolution of passive streaker setups, in particular the effect of an energy chirp on the measured slice energy spread.
	Downstream of the hard X-ray SASE2 beamline at the European XFEL, these measurements are enabled by a single-plate non-movable passive wakefield streaker, essentially a rectangular corrugated plate placed inside a vacuum chamber.
	We show measurements with a time resolution down to a few femtoseconds, and an energy resolution down to a few MeVs.
\end{abstract}

\maketitle

\section{Introduction}
\label{sec:intro}

Electron beam longitudinal phase space (LPS) measurements performed after Free-Electron Laser (FEL)~\cite{Pellegrini2016,Huang2021} undulator beamlines provide information that is highly relevant for FEL operation and machine development, in particular the beam current profile, the energy chirp (the change of the slice mean energy along the bunch), and the slice energy spread.
High beam currents are important for FEL performance and efficiency.
Moreover, knowledge of the current profile is required to calculate the undulator wakefields.
Knowledge of the energy chirp and the undulator wakefields allows to predict the optimal undulator taper profile~\cite{Tomin2024}.
A strong chirp imposes a relation between time and central frequency of the FEL pulse, interesting for some applications~\cite{Fadini2020}, and increases the FEL bandwidth.
While the vast majority of users at FEL facilities request a minimal bandwidth, others benefit from maximal bandwidth~\cite{Nass2021}.
A small slice energy spread is fundamental for high-gain FELs~\cite{Kondratenko1980,Bonifacio1984}, and measurements are helpful for optimizations such as finding the optimal laser heater~\cite{Saldin2004} working point.
Additionally, LPS measurements are the basis for indirect FEL pulse duration and power profile measurements~\cite{Ding2011,Behrens2014,Dijkstal2022}, which are among the most important FEL parameters.
In particular, many users of X-ray FEL facilities request short pulses~\cite{Chapman2014,Young2018,Chapman2019}.

The LPS is typically measured with transversely deflecting rf structures (TDS) that impose a linear correlation between time and transverse beam coordinates~\cite{Emma2000,Ding2011,Behrens2014,Floettmann2014}, and a transverse profile imager~\cite{Wiebers2013} (beam screen) at a dispersive location.
In recent years passive wakefield streakers (PS) have emerged as an alternative to TDS~\cite{Bettoni2016,Seok2018,Dijkstal2022,Tomin2022}, and are now employed for time-resolved measurements of multi-GeV electron beams at the SwissFEL~\cite{Prat2020b} Aramis beamline and the European XFEL~\cite{Decking2020} (EuXFEL) SASE2 beamline.
In a PS the bunch receives a time-dependent transverse kick similar to a TDS.
There are two PS variants, which generate wakefields either from corrugated~\cite{Novokhatski2015} or dielectric~\cite{Bettoni2016} surfaces.
Compared to a TDS, a PS has far lower construction and operating costs.
Furthermore, wakefield streaking is inherently synchronized to the beam arrival time.
The main disadvantage of PS is the nonlinearity of wakefield streaking, which causes a varying time resolution along the bunch.
In particular, the head of the beam is not resolved.
Moreover, complex analysis algorithms are required.

LPS diagnostics ideally have time and energy resolutions better than the values to be measured.
The final bunch durations at hard X-ray FEL user facilities are on the order of few tens of fs, while the FEL pulses can be shorter: sub-fs durations have been achieved (e.g.,~\cite{Huang2017,Trebushinin2023}).
At different FEL facilities, the beam slice energy spread at the beginning of the undulator section was measured at the 100~keV level~\cite{Penco2017} (FERMI~\cite{Allaria2012}), and the MeV level~\cite{Prat2024} (SwissFEL).
The high-gain FEL instability inside the undulator section can increase the slice energy spread to the 10~MeV level~\cite{Behrens2014,Dijkstal2022b}.
Therefore, LPS measurements of the non-lasing beam have higher requirements on the energy resolution compared to measurements of the lasing beam at optimized performance.
Previous publications on PS diagnostics focus on lattice design for an optimal time resolution~\cite{Craievich2017}, current profile measurements~\cite{Bettoni2016,Seok2018} and FEL power profile measurements~\cite{Dijkstal2022}, less so on standalone LPS measurements.
We show in this paper that the energy resolution in LPS measurements based on electron streaking is slice-dependent, and that the energy chirp has an impact on the observed slice energy spread.
The varying time resolution of PS diagnostics exacerbates this effect.
Here we provide the theoretical understanding necessary to interpret LPS measurements.

Double-sided movable corrugated PS are used for diagnostics purposes at SwissFEL~\cite{Dijkstal2023} (and at relatively low beam energy at the PAL-XFEL injector test facility~\cite{Han2015,Seok2018}).
The PS at EuXFEL has a simpler design with a single, non-movable corrugated plate, which is advantageous in terms of vacuum safety and cost-efficiency.
It has previously been successfully employed to observe qualitative features of the FEL pulse~\cite{Tomin2022}, and for slice energy spread measurements without time calibration~\cite{Tomin2023}.
In this paper, we show its first time-resolved measurements of important properties such as the electron beam energy chirp and the FEL pulse duration.

An analytical wakefield model~\cite{Bane2016,Bane2016a} is the basis for our analysis.
It contains the transverse dipole and quadrupole components and the longitudinal component up to first order.
The transverse dipole component deflects the beam towards the structure.
The transverse quadrupole component has a defocusing effect in streaking direction, and a focusing effect in the other transverse direction.
The longitudinal component causes a slice energy loss.
All effects are weakest at the head and grow stronger along the bunch.

In this paper we first describe the experimental setup at EuXFEL in Sec.~\ref{sec:experimental_setup}.
Then we present the expected time and slice energy resolution of a PS diagnostics setup in Sec.~\ref{sec:eff_meas_res}.
Afterwards we discuss LPS and FEL power profile measurements in Sec.~\ref{sec:measurements}.

\section{Experimental setup}
\label{sec:experimental_setup}

\begin{figure}
	\begin{center}
	\includegraphics[width=\linewidth]{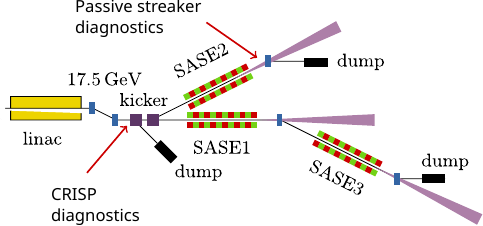}
	\end{center}
	\caption{Layout of EuXFEL after the final compression stage (taken from~\cite{Lockmann2020}, modified).
	}
	\label{fig:euxfel}
\end{figure}

The European XFEL is a high repetition rate X-ray FEL user facility powered by a superconducting linac~\cite{Decking2020}.
It accelerates 10 macropulses per second, each of which contains up to 2700 electron bunches.
The final beam energy is up to 17.5~GeV, the nominal bunch charge is 250~pC, and the normalized transverse emittance is less than 0.6~\textmu{}m.
The bunches are distributed among three undulator beamlines, where they generate intense X-ray FEL radiation with tunable photon energies between approximately 0.25 and 25~keV.
The FEL pulses are used in the experimental end stations for research of ultra-fast phenomena in matter at the atomic scale.
Beam optics are regularly measured and matched in the injector section~\cite{Meykopff2018}.
The CRISP THz spectrometer non-invasively measures the current profile at the end of the linac section~\cite{Lockmann2020}, with a time resolution that can generally be estimated at 8~fs, and averages over 16 bunches.
X-ray gas monitors measure the FEL pulse energy with an uncertainty of less than 10\%~\cite{Maltezopoulos2019}.
Figure~\ref{fig:euxfel} shows the relevant part of the EuXFEL layout.

\begin{figure}
	\begin{center}
	\includegraphics[width=\linewidth]{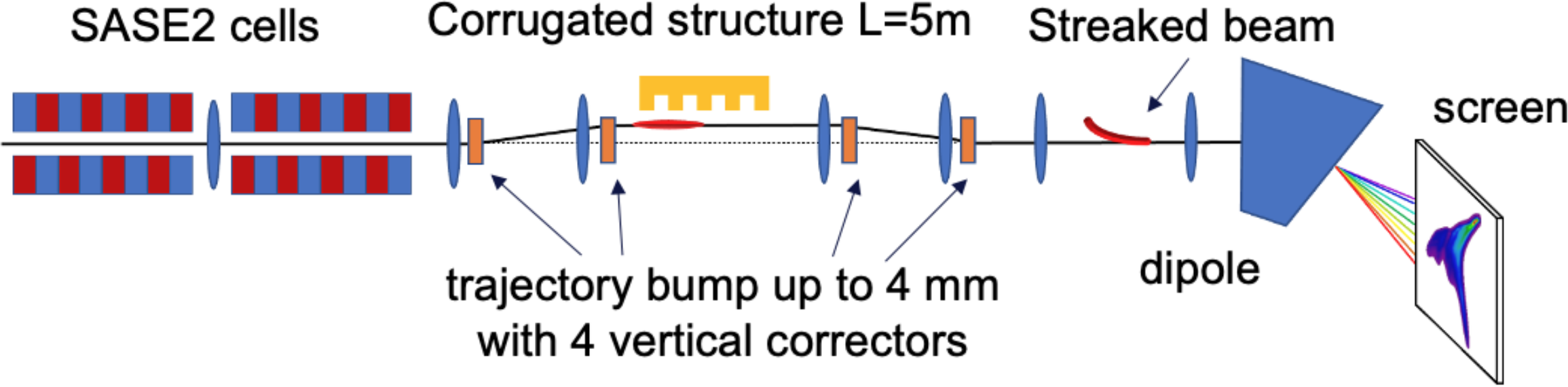}
	\\
	\includegraphics[width=0.4\linewidth]{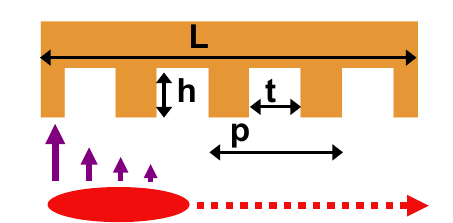}
	\end{center}
	\caption{Top: layout of the diagnostic beamline for LPS measurements after the SASE2 undulator beamline.
Bottom: geometry of the corrugated plate.
The growing wakefield fores along the bunch are indicated by purple arrows.
	}
	\label{fig:setup}
\end{figure}

The EuXFEL passive streaker, designed at DESY, is essentially a vacuum chamber that contains a 5~m long rectangular corrugated plate.
Steering dipoles can move the beam closer to the plate and thereby control the streaking magnitude.
A larger orbit bump results in a smaller distance $d$ between beam and plate, and in stronger wakefields.
The first dipole in the arc after SASE2 serves as an energy spectrometer in the horizontal plane.
Several measurement optics with optimized time resolution and with different dispersion values at the beam screen in the arc can be set~\cite{Tomin2022}.
In general a beam screen can only resolve one bunch per macropulse.
With the PS in operation, beam losses further limit the beam repetition rate to one bunch in every third macropulse.
For measurements of the non-lasing beam we use a steering dipole to introduce horizontal betatron oscillations within SASE2, and thereby disable the FEL process.
Figure~\ref{fig:setup} shows the PS diagnostics setup after the SASE2 beamline, and a schematic of the corrugation geometry.
Table~\ref{tab:parameters} contains important parameters of the diagnostics setup.

\begin{table}
	\caption{The corrugation parameters according to the sketch in Fig.~\ref{fig:setup} bottom left, and the relevant beam optics and transport lattice parameters for the measurements shown in this paper~\cite{Tomin2022}.}
	\centering
	\begin{tabular}{lll}
		\hline\hline
		\multirow{5}{*}{Corrugation geometry} & Depth, $h$ & 0.5~mm \\
		 & Gap, $t$ & 0.25~mm \\
		 & Period, $p$ & 0.5~mm \\
		 & Width, $w$ & 12~mm \\
		 & Length, $L$ & 5~m \\\hline
		 \multirow{2}{*}{Beam optics at PS} & $\beta_x$, $\alpha_x$ & 34.5~m, 1.0 \\
		 & $\beta_y$, $\alpha_y$ & 54.2~m, -1.9\\\hline
		 \multirow{3}{*}{Transport lattice} & Dispersion & 30~cm\\
		 & $R_{34}$ & -40~m\\
		 & $\Delta \mu_y$ & $281^\circ$\\\hline
		 Beam optics at screen & $\beta_x$, $\beta_y$ & 4.9~m, 30~m\\
		 \hline\hline
	\end{tabular}
	\label{tab:parameters}
\end{table}

\section{Measurement resolution}
\label{sec:eff_meas_res}

In the following we first recall the time resolution theory for passive streakers, and find an expression for the time-dependent energy resolution for LPS measurements based on electron streaking.
Then we confirm these results with ideal beam macroparticle simulations.

\subsection{Expressions for time and energy resolution}
\label{sec:time_energy_res}

While we want to measure the time and energy ($t$ and $E$) distribution in an LPS measurement, we can only see the transverse ($y$ and $x$) distribution on a beam screen.
For this reason we impose correlations between $t$ and $y$, and between $x$ and $E$.
Yet as a consequence of the natural beam size and the complex structure wakefield effects, particles with different $t$ or $E$ coordinates can have the same $y$ or $x$ coordinates at the screen, and thus the resolving power of the diagnostics is limited.
The time and energy resolution specify the smallest temporal and energetic features of the beam that can be measured.
Lower resolution values are preferable.
We emphasize that the main cause of the resolution is beam dynamics, not the camera pixel size or the finite screen resolution $\sigma_R$.
In our case $\sigma_R$ is better than 10~\textmu{}m~\cite{Wiebers2013}, similar to or less than the natural rms beam sizes at the screen position.
Therefore, even ideal hardware ($\sigma_R=0$) and infinitesimally fine image slicing would not significantly improve the measurement resolution.

The rms time resolution $r_t$ of a PS measurement setup with transverse dipole and quadrupole wakefields components is~\cite{Craievich2017}:
\begin{align}
	r_t(t) = \frac{\sigma_y(t)}{\left|\mathrm{d}y(t)/\mathrm{d}t\right|},
	\label{eq:time_res}
\end{align}
with the time-resolved slice beam size at the screen in streaking direction $\sigma_y(t)$, and the streaking term at the screen $\mathrm{d}y(t)/\mathrm{d}t$.
The screen resolution $\sigma_R$ adds in quadrature to the beam size, but we omit it for clarity.
The streaking term is zero at the head of the beam, and progressively increases along the bunch.
Therefore, $r_t$ diverges at the head.
The defocusing quadrupole wakefields are negligible at the head, but increase towards the tail at a faster rate than the dipole wakefields responsible for the streaking.
Reference~\cite{Craievich2017} contains expressions of $\sigma_y$ as a function of transverse emittance and Twiss parameters at the PS, of quadrupole wakefield strength, and of the transport lattice between structure and screen.

In the following we consider the rms energy resolution $r_E$ and the interplay with the time resolution.
The energy resolution $r_E$ of a dipole spectrometer setup is typically given as~\cite{Ding2011}
\begin{align}
	r_E = E_0\frac{\sigma_{x0}}{|D|},
	\label{eq:energy_res}
\end{align}
with the average beam energy $E_0$, the dispersion $D$, and the slice beam size at the screen $\sigma_{x0}$ (in absence of dispersion).
For TDS setups it can be reasonable to assume that $\sigma_{x0}$ is the same for each slice.
Conversely, the quadrupole wakefields in a PS setup act as a quadrupole magnet with time-dependent strength and give a slice optics mismatch along the bunch~\cite{Qin2017}.
However, we find that Eq.~(\ref{eq:energy_res}) does not explain the large slice energy spread we measure with a PS:
\begin{align}
	\sigma_{Em}(t) = E_0 \frac{\sigma_x(t)}{|D|},
	\label{eq:espread_meas}
\end{align}
with $\sigma_x(t)$ the dispersive beam size measured on the screen.
We introduce the slice energy resolution $r^*_E$ to quantify the difference between measured ($\sigma_{Em}$) and true initial ($\sigma_E$) energy spread:
\begin{align}
	r^*_E(t) \equiv \sqrt{\sigma^2_{Em}(t) - \sigma^2_E(t)},
	\label{eq:effective_eres}
\end{align}
and identify two effects that increase $\sigma_{Em}$ in addition to the natural beam size effect.

First, any additional slice energy spread $\Delta \sigma_E$ induced by the PS (if placed before the energy spectrometer) contributes to $\sigma_{Em}$.
The corresponding analytical expression is absent from Refs.~\cite{Bane2016,Bane2016a}, but is easily calculated from the information therein (appendix~\ref{sec:long_wake}).
We note that also TDS increase the slice energy spread, which is corrected in high-resolution energy spread measurements~\cite{Floettmann2014,Ratner2015,Prat2020c,Prat2024}.

Second, due to the finite time resolution, particles from different time slices end up at the same $y$ position on the screen.
If these time slices have different mean energies, the measured energy spread (of all particles with the same $y$ coordinate) is larger than the true energy spread (of all particles with the same $t$ coordinate).
We call this effect a spillover of time to energy resolution.
Consistent with this description, in measurements of beams with quadratic energy chirp the minimum slice energy spread is found at the time slice with zero chirp~\cite{Prat2020c}.
In appendix~\ref{sec:effective_long} we analytically verify the effect for a simplified case.

\begin{figure*}
    \centering
	\includegraphics{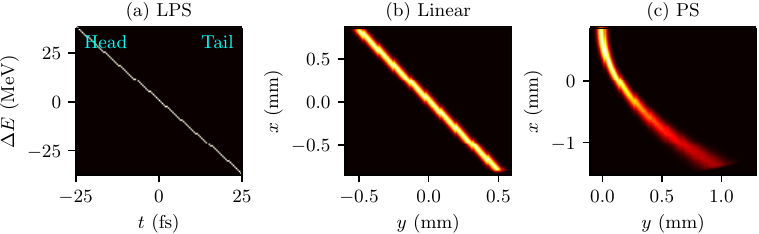}
    \caption{Ideal beam simulations of LPS measurements based on electron streaking.
(a): initially generated beam with zero energy spread.
(b-c): simulated transverse distribution at the screen for two cases: linear streaking (b), and wakefield streaking (c).
We display the $y$ coordinate in horizontal direction for consistency with the conventional display of LPS measurements.
	}
    \label{fig:sim_lps}
\end{figure*}

Neglecting correlations between transverse and energy coordinates imposed by higher-order wakefields, and assuming a slowly-varying current profile, the total slice energy resolution $r^*_E(t)$ is the quadratic sum of the three effects:
\begin{align}
	r^*_E(t) = \sqrt{ \left(\frac{E_0\sigma_{x0}(t)}{D}\right)^2 + \Delta\sigma_E^2(t) + \left(r_t(t)\frac{\mathrm{d}E(t)}{\mathrm{d}t}\right)^2}.
	\label{eq:energy_res3}
\end{align}
The first term (proportional to $\sigma_{x0}$) varies slightly along the beam due to the focusing quadrupole wakefield effect.
The second term ($\Delta \sigma_E$) steadily increases along the bunch.
The third term is proportional to the energy chirp $\mathrm{d}E(t)/\mathrm{d}t$ and can have any behavior.
We note that in most situations where the LPS is measured, it would be nonsensical to beforehand alter the LPS for minimal chirp and optimal $r^*_E$.
Only in some cases this is feasible, such as for the measurements of the quantum diffusion effect previously performed at EuXFEL with the PS~\cite{Tomin2023}.

\subsection{Simulated LPS measurement}
\label{sec:sim_res}

We illustrate the resolution effects described above by simulating LPS measurements based on electron streaking.
We assume that the current profile and the distance between beam and corrugated plate, and thus the PS wakefield potentials, are known.
First we generate a 6D macroparticle distribution describing the incoming electron beam.
For simplicity, we choose a rectangular 5~kA current profile.
Inspired by measurements shown later, we set the linear energy chirp to -1.5~MeV/fs.
Moreover, the initial beam has zero slice energy spread, thus we have the condition $\sigma_{Em} = r^*_E$.
Other parameters are 14~GeV mean beam energy, 250~pC bunch charge, the transverse optics at the PS location listed in Tab.~\ref{tab:parameters}, and a normalized transverse emittance of 0.6~\textmu{}m.
For the time streaking effect we consider two variants.
First, linear streaking, essentially an ideal TDS setup neglecting its side effects on the beam energy~\cite{Floettmann2014} (\textit{Linear}).
Second, the PS case with all wakefield components enabled (\textit{PS}).
In both cases, we choose the streaking strength such that the total extent of the beam along the streaked coordinate $y$ is approximately 1~mm.
For the \textit{PS} case we use $d=600$~\textmu{}m.
Finally, we track the particle distribution to the screen using the linear transport optics described in Tab.~\ref{tab:parameters}.
Figure~\ref{fig:sim_lps} shows the initial distribution in $t$ and $E$ coordinates (a), and the distributions in $y$ and $x$ coordinates at the screen for the two cases (b-c).

\begin{figure}
    \centering
	\includegraphics{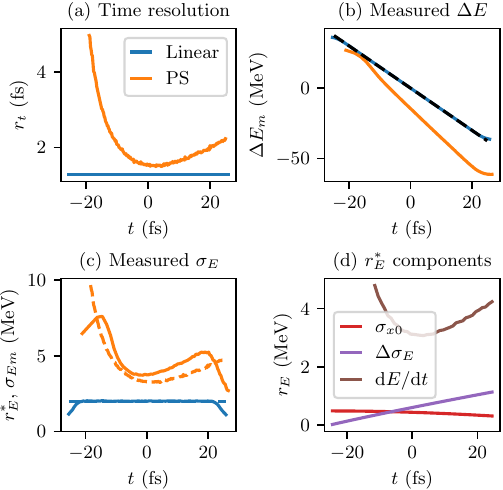}
    \caption{
Slice mean energy and energy spread analysis of the simulated LPS of Fig.~\ref{fig:sim_lps}.
(a): time resolution of the measurements.
(b): reconstructed slice mean energy $\Delta E_m$ (colors), and true $\Delta E_m$ (dashed black).
(c): reconstructed slice energy spread $\sigma_{Em}$ in solid, and slice energy resolution $r^*_E$ in dashed and matching colors.
Colors in (b-c) refer to the legend of plot (a).
(d): the three components of $r^*_E$ as in Eq.~(\ref{eq:energy_res3}) for the \textit{PS} case.
    }
    \label{fig:sim_meas}
\end{figure}

In the analysis we first calculate the time resolution by directly evaluating Eq.~(\ref{eq:time_res}) from the $y$ macroparticle coordinates at the screen, and the initial $t$ coordinates.
We note that the reconstruction of the time coordinates refers to the beam at the PS and not at the screen, even though longitudinal dispersion introduced by the spectrometer dipole alters the current profile between these two positions.
As expected, we obtain a constant $r_t$ for the \textit{Linear} case, and a diverging $r_t$ at the head of the beam for the \textit{PS} case (Fig.~\ref{fig:sim_meas}a).
For \textit{PS}, we obtain a minimal $r_t$ below 2~fs at the center.

Next we imitate a measurement of the slice mean energy $E_m$ and the slice energy spread $\sigma_{Em}$ by projecting the 6D particle distribution to just $x$ and $y$, and slicing it along $y$.
In the trivial linear streaking case we obtain the $t$ axis from a division of the $y$ axis by the streaking term $\Delta y/\Delta t$.
In the \textit{PS} case we retrieve the $t$ axis from the transverse dipole wakefield potential.
We calculate $E_m$ and $\sigma_{Em}$ from a center of mass and rms calculation and evaluate $r^*_E$.
Figure~\ref{fig:sim_meas} (b-d) displays the results of our analysis.
Plot (b) shows the difference between the true (black) and measured (colors) slice mean energy deviation.
There is excellent agreement in the \textit{Linear} case, whereas the \textit{PS} case includes the energy loss from the longitudinal wakefields and thus gives a different result.
This underscores the importance of correcting the slice mean energy loss from the longitudinal structure wakefields.
Knowledge of the current profile and of $d$ is sufficient to calculate the main effect.
We find the energy loss correction term described in appendix~\ref{sec:long_wake} to be negligible.

In Fig.~\ref{fig:sim_meas} (c) we show the numerically evaluated slice energy spread $\sigma_{Em}$ (solid) compared to the prediction $r^*_E$ (dashed).
For the \textit{Linear} case we find exact agreement.
For the \textit{PS} case there is good agreement, with a minimal $r^*_E$ between 3.5 and 4~MeV at the center of the bunch.
However, the actual measured energy spread is slightly underestimated by $r^*_E$.
Plot (d) displays the three components of $r^*_E$ for the \textit{PS} case (see Eq.~\ref{eq:energy_res3}).
The contribution of the natural beam size ($\sigma_{x0}$) is on the order of 0.5~MeV, and decreases towards the tail of the beam due to the focusing quadrupole wakefield effect.
The energy spread induced by the PS ($\Delta\sigma_E$) has an approximately linear growth along the bunch, and reaches up to 1~MeV at the tail.
The energy chirp term is the by far largest contribution, with a minimum of around 3~MeV at the beam center.
The simulations show that the expected slice energy resolution of the energy-chirped beam is on the order of several MeV, higher than the energy spread expected for the non-lasing beam at an FEL facility, and less than what is expected for the lasing beam.

In appendix~\ref{sec:benchmark} we compare the macroparticle tracking functionality of our analysis software~\cite{Dijkstal2022b} and the OCELOT code~\cite{Agapov2014}.
We find good agreement between the codes for the values of $d$ we typically set, which includes the measurements and simulations shown in this paper.
Only for smaller $d$, the more precise but also more time-consuming wakefield implementation of OCELOT is preferable.

\section{Measurements at EuXFEL}
\label{sec:measurements}

In the following pages we analyze various measurements performed with the PS after SASE2.
We start with a single LPS measurement (with the FEL disabled) to illustrate the procedure, in particular the time calibration.
Afterwards we analyze three LPS measurements of the same beam conditions under variation of $d$.
Finally, we also demonstrate an example FEL power profile measurement.

\subsection{Time calibration and LPS measurement}
\label{sec:time_calib}

In our analysis procedures~\cite{Dijkstal2022,Dijkstal2022b} we first determine $d$ through an iterative process.
In essence, we record the current profile measured by CRISP and the vertical beam center of mass on the screen before and after the streaking conditions are set.
In PS macroparticle tracking simulations using the measured current profile, we vary $d$ until we match the center of mass shift recorded earlier.
Then we calculate the impact of the wakefield kick at the position of the beam screen, using the transverse dipole wakefield potential and the $R_{34}$ beam transport matrix element.
This lets us convert the vertical image axis to time.
We ensure that the calibration is reasonable by comparing the shape and rms duration of the current profile as reconstructed from the transverse beam profile (using the algorithm from Ref.~\cite{Dijkstal2022}) to the CRISP measurement.
Finally, we convert the horizontal image axis to energy with the known dispersion value.
We note that the assumption of transverse beam optics does not affect the simulated center of mass deflection of the beam, and thus also not the distance calibration procedure.

\begin{figure}
	\centering
	\includegraphics{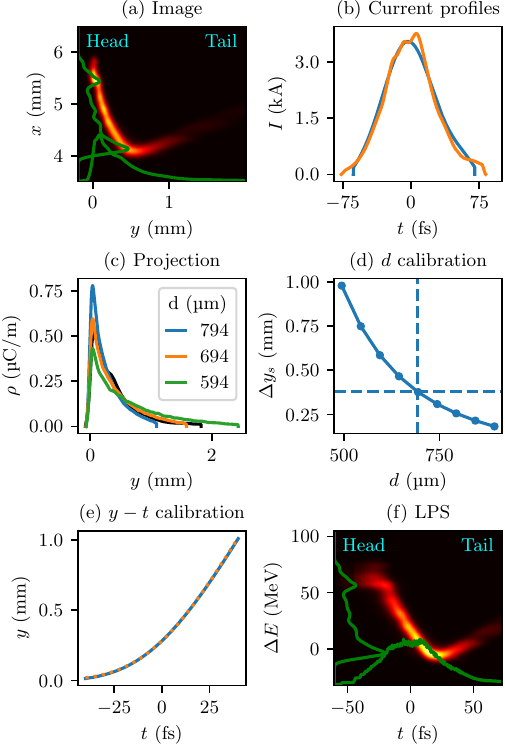}
	\caption{LPS measurement including distance ($d$) calibration procedure.
(a): measured transverse projection of the streaked beam, again with the $y$ coordinate in horizontal direction for consistency.
(b): CRISP (blue) and reconstructed (orange) current profiles.
(c): measured (black) and simulated (colors) streaked projections at the screen assuming three different $d$.
(d): simulated transverse center of mass shift.
The dashed horizontal line indicates the measured shift.
The dashed vertical line indicates the calibrated $d$ of 694~\textmu{}m.
(e): time calibration curves, using the current profiles from (b) as input (same colors).
(f): reconstructed LPS.
	}
	\label{fig:forward}
\end{figure}

The calibration and LPS measurement procedure is demonstrated in Fig.~\ref{fig:forward}.
Plots (a-b) show the recorded screen image and the CRISP profile, which serve as input to the calibration procedure.
Plots (c-d) show the outcome of the particle tracking: the transverse projections on the screen (c), and the center of mass shift (d).
We also find good agreement between the reconstructed and CRISP current profiles (b).
The calibration curves obtained from the dipole wake potentials of the two current profiles are virtually identical (e).
We use the calibration curve from the CRISP profile to convert the raw image to LPS (f).

\subsection{Slice energy properties measured under variation of streaking strength}
\label{sec:distance_scan}

Earlier we discussed the influence of the time resolution $r_t$ on the slice energy resolution $r^*_E$, and the influence of the longitudinal wakefields on the slice mean energy.
Both effects have a strong dependency on $d$.
In the following we analyze LPS measurements taken with different beam trajectories inside the PS.

\begin{figure}
	\centering
	\includegraphics{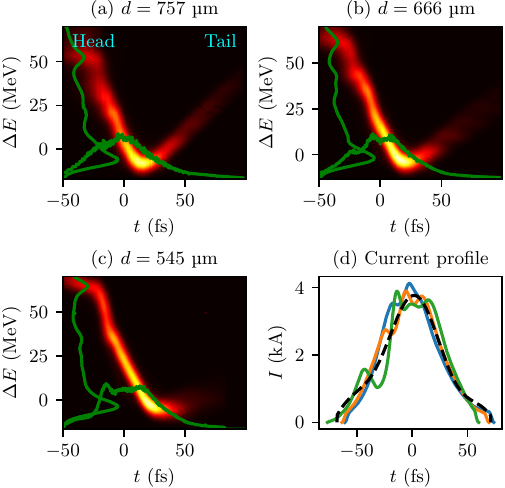}
	\caption{(a-c): three LPS measurements acquired under the same beam conditions but varying beam trajectory inside the PS.
The calibrated $d$ are listed in the titles.
(d): reconstructed current profiles from the PS (colors) and from CRISP (dashed).
	}
	\label{fig:multi_lps}
\end{figure}

Figure~\ref{fig:multi_lps} (a-c) shows three consecutive LPS measurements, between each of which we decreased the $d$ set value by 100~\textmu{}m.
The calibrated $d$ changes by 79 and 121~\textmu{}m, giving differences between result and expectation of $\pm$21~\textmu{}m which we attribute to errors in the set trajectory.
For a consistency check we compare CRISP and reconstructed current profiles, and find good agreement in each case (d).
The effect of the microbunching instability~\cite{Ratner2015} is clearly visible in the single-shot PS measurements, but not in the CRISP measurement that is averaged over several shots.

\begin{figure}
    \centering
    \includegraphics{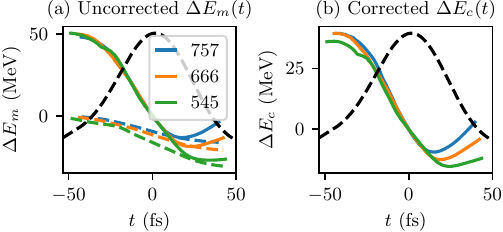}
    \caption{Slice mean energy analysis for the LPS measured at different $d$ as shown in Fig.~\ref{fig:multi_lps}.
(a): raw slice mean energy $\Delta E_m$ (solid). Estimated longitudinal wakefield contribution to $\Delta E_m$.
The legend indicates $d$ in units of \textmu{}m.
(b): Corrected slice mean energy $\Delta E_c$.
In both plots, the CRISP current profile is overlaid in arbitrary units (dashed black), and the relative energy is shifted to zero at $t=0$.
    }
    \label{fig:meanE}
\end{figure}

In Fig.~\ref{fig:meanE} we analyze the slice mean energy.
We limit the analysis to parts of the beam with a current over 1~kA, such that the results are less affected by noise.
Plot (a) shows the uncorrected $\Delta E_m$.
At the head and center, for all $d$ we measure a negative energy chirp of -1.5~MeV/fs.
Towards the tail, the chirp decreases and turns to positive values, although there are pronounced differences of up to 15 MeV between the three measurements.
We calculate the expected wakefield energy loss based on the CRISP current profile, and display the corrected slice energy values $\Delta E_c$ in (b).
At the head and center, the energy chirp is reduced to about -1.2~MeV/fs.
At the tail, differences between the measurements persist but are smaller in magnitude by about a factor two.
The analysis indicates that we underestimate the slice energy loss imposed by the PS at the tail, and thus cannot accurately measure the energy chirp there.

\begin{figure}
    \centering
    \includegraphics{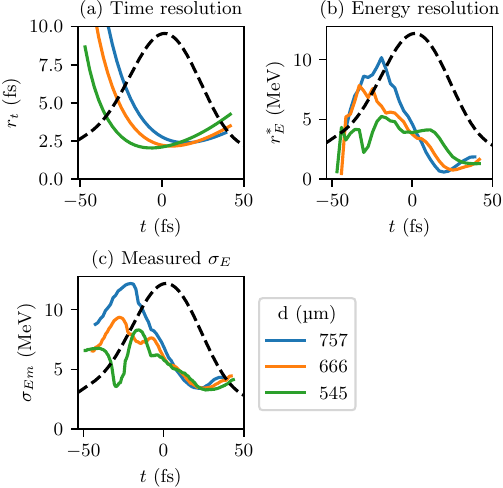}
    \caption{Slice energy spread and resolution analysis for the LPS measured at different $d$ as shown in Fig.~\ref{fig:multi_lps}.
(a-b): calculated time and slice energy resolution.
(c): measured slice energy spread.
In all plots, the CRISP current profile is overlaid in arbitrary units (dashed black).
    }
    \label{fig:spreadE}
\end{figure}

In Fig.~\ref{fig:spreadE} we analyze the measurement resolution and the measured energy spread.
We begin with $r_t$, which is evaluated assuming design transverse optics (a).
The minimum values of $r_t$ are close to 2.5~fs for each $d$.
Plot (b) shows $r^*_E$ which we calculated using $\mathrm{d}E/\mathrm{d}t$ extracted from $E_m$ shown in Fig.~\ref{fig:meanE} (a).
Finally, we display $\sigma_{Em}$ (c).
The large differences between the three measurements indicate strong resolution effects, and are expected considering the impact of $d$ on $r_t$.
Consistent with the resolution theory, the minimum energy spread (about 3.5~MeV in all cases) is observed for the slices without uncorrected energy chirp.
The time coordinate of this slice changes with $d$ because of the additional energy chirp imposed by the PS.
The calculated $r^*_E$ are generally similar in shape to the measured $\sigma_{Em}$.
However, an evaluation of the true slice beam energy spread through subtraction of $r^*_E$ (see Eq.~\ref{eq:effective_eres}) would still yield implausibly large values, indicating that we underestimate $r^*_E$.

\subsection{FEL power profile measurement}
\label{sec:fel_meas}

For an FEL power profile measurement we combine the slice properties acquired at lasing-enabled and lasing-disabled beam orbit conditions in the undulator.
The FEL power profile can be reconstructed using two methods~\cite{Behrens2014}.
First through the slice mean energy loss caused by the FEL ($P_\Delta$), and second through the increase in slice energy spread $\sigma_E$ caused by the FEL ($P_\sigma$), with
\begin{align}
P_\sigma(t) \propto I^{2/3}(t)(\sigma^2_{E,\mathrm{on}}(t) - \sigma^2_{E,\mathrm{off}}(t)).
\end{align}
We typically acquire 20 LPS measurements for each configuration, then use the average $\sigma_E$ values to calculate the average $P_\sigma$, which we then normalize through the X-ray gas monitor pulse energy measurement.
Afterwards we use the same normalization factor in combination with the single-shot $\sigma_{E,\mathrm{on}}$ and the average $\sigma_{E,\mathrm{off}}$ to obtain single-shot $P_\sigma$.
An important feature of $P_\sigma$ is that the impact of $r^*_E$ cancels in case the vertical beam trajectory inside the PS is maintained for the two FEL configurations.
We do not correct the LPS measurements for the incoming mean beam energy jitter (which would be possible), and neither for the energy chirp jitter (which would be difficult).
Both jitters have a large impact on $P_\Delta$ and a small impact on $P_\sigma$.
Thus we only show $P_\sigma$ in the following.

\begin{figure}
	\centering
	\includegraphics{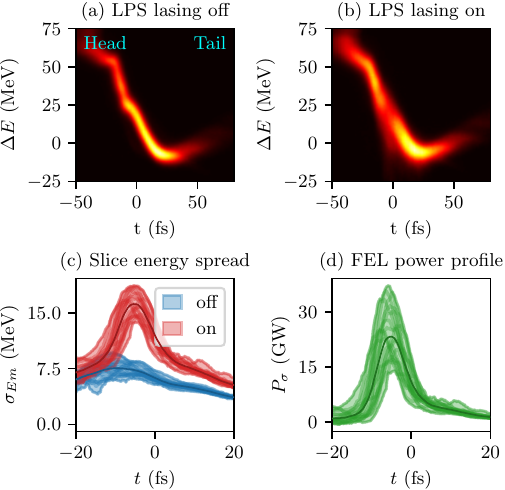}
	\caption{LPS measurements at lasing-disabled (a) and lasing-enabled (b) undulator configurations are combined for a FEL power profile measurement.
(c) shows the $\sigma_{Em}$ values for the two conditions, and (d) shows the FEL power profiles reconstructed from the increased energy spread when the FEL is enabled.
The lightly colored lines show single-shot values, and the darkly colored lines the average value of 20 shots.
	}
	\label{fig:lasing}
\end{figure}

We performed an FEL power profile measurement with the same machine conditions as for the measurements presented earlier.
The undulators were tuned to a resonant photon energy of 9~keV, and the average FEL pulse energy measured with the X-ray gas monitor was 300~\textmu{}J.
The average calibrated $d$ is 620~\textmu{}m.

Figure~\ref{fig:lasing} shows our results.
The energy spread blow-up caused by the FEL is clearly visible in the example LPS (a-b).
Fig.~\ref{fig:lasing}~(c) show the measured slice energy spread values, which results in the power profiles in (d).
Within the lasing part of the beam, $\sigma_{E,\mathrm{off}}$ (also an upper limit of $r^*_E$) is up to 7.5~MeV, while $\sigma_{E,\mathrm{on}}$ is up to 15~MeV, indicating a slice energy resolution sufficient for the FEL power profile measurement.
The average full width at half maximum (FWHM) duration $\tau$ of the single-shot $P_\sigma$ is 8.6~fs, with an rms shot-to-shot variation of 1.5~fs.

A systematic error term for $\tau$ is the time resolution, which leads to an overestimation of the FEL pulse duration.
For the lasing part as in this power profile measurement (between approximately -10 and 5~fs), we assume an $r_t$ around 2.5~fs (see Fig.~\ref{fig:spreadE}~b).
From the subtraction in quadrature of $r_t$ from the measured values of $\tau$ (using a factor 2.335 to convert rms values to FWHM) we estimate a true FWHM bunch duration of 6.3~fs, meaning that $r_t$ would have a relative impact of around 35\%.
In most measurements we observe a much longer FEL pulse duration, in which case $r_t$ only has negligible effect.

\section{Conclusion}

We presented time-resolved diagnostics at the European XFEL with a corrugated passive wakefield streaker.
Compared to movable double-sided PS, our non-movable single-sided PS has an even bigger advantage in cost-effectiveness over TDS diagnostics.
We exploit the non-invasive CRISP diagnostics~\cite{Lockmann2020} for a reliable and consistent calibration of the beam trajectory.

To interpret the LPS measurements of the energy-chirped beam after the SASE2 beamline, we developed a slice energy resolution theory for electron streaking measurement setups, complementing an earlier study on the time resolution for PS diagnostics~\cite{Craievich2017}.
We find in simulations and measurements that the spillover of time into energy resolution is a major effect in our case.
We showed LPS measurements with a time resolution down to few femtoseconds, and a slice energy resolution down to few MeV for time slices with small local energy chirp.
We showed an example FEL power profile measurement with an average FWHM photon pulse duration below 10~fs.

We also indicated which properties of the LPS our PS diagnostics cannot resolve.
In particular, we cannot measure conclusively the mean slice energy of the beam tail, which we attribute to an insufficient understanding of the slice energy loss induced by the PS.
We also cannot measure the slice energy spread of the nonlasing beam with sufficient resolution.
For this purpose we suggest to employ the optical klystron effect instead~\cite{Penco2017,Prat2024}.

The data and analysis software used for the results shown in this paper are available for download~\cite{Dijkstal2024}.

\begin{acknowledgments}
We thank M.\ K.\ Czwallina, W.\ Decking, L.\ Fröhlich, N.\ Golubeva, A.\ Leuschner, D.\ Lipka, N.\ Lockmann, A.\ Novokshonov, and T.\ Wohlenberg for their support in operating and improving the passive streaker diagnostics.
We acknowledge all technical groups at DESY involved in the PS design and installation.
We thank K.\ Bane, M.\ Guetg, S.\ Reiche, and I.\ Zagorodnov for helpful discussions.
\end{acknowledgments}

\appendix

\section*{Appendix}

\section{Slice energy spread induced by longitudinal wakefields}
\label{sec:long_wake}

In the wakefield model~\cite{Bane2016,Bane2016a}, the transverse effects are described by dipole ($w_{yd}$) and quadrupole ($w_{yq}$) wakefield components:
\begin{align}
	&w_y(s, y) = w_{yd}(s) + y\ w_{yq}(s)
	\label{eq:spw1}
	\\
	&w_x(s, x) = - x\ w_{yq}(s),
	\label{eq:spw2}
\end{align}
and the longitudinal effect is described by $w_l$.
All wakefield terms depend on the longitudinal distance $s$ between drive and test particle.
The quadrupole effect depends on the individual transverse test particle coordinates $x$ and $y$ relative to the beam centroid, while the other two effects do not.
The energy spread increase imposed by the longitudinal wakefield component~\cite{Bane2016b} is not explicitly calculated in Refs.~\cite{Bane2016,Bane2016a}, but we evaluate it here through application of the Panofsky-Wenzel theorem~\cite{Dohlus2017}:
\begin{align}
	\frac{\partial}{\partial s} w_x = -\frac{\partial}{\partial x} w_l \ ;\quad
	\frac{\partial}{\partial s} w_y = -\frac{\partial}{\partial y} w_l,
	\label{eq:panofsky_wenzel}
\end{align}
where $x$ and $y$ refer to the transverse test particle coordinates, and $s$ is the longitudinal coordinate (directed opposite the main direction of movement).
The left hand sides of the two equations above exist, whereas the available expressions for $w_l$ have no $x$ or $y$ dependencies.
Inserting Eqs.~(\ref{eq:spw1}) and (\ref{eq:spw2}), we obtain one correction term from each $w_{yd}$ and $w_{yq}
$\begin{align}
	&-\int \mathrm{d}y\ \frac{\partial}{\partial s} w_{yd} = -y \frac{\partial}{\partial s} w_{yd}
	\\
	&\int \mathrm{d}x\ x\frac{\partial}{\partial s} w_{yq} - \int \mathrm{d}y\ y\frac{\partial}{\partial s} w_{yq}
	= \frac{x^2-y^2}{2} \frac{\partial}{\partial s} w_{yq},
\end{align}
which we use to define the corrected longitudinal wake function $w^*_l$:
\begin{align}
	&w^*_l \equiv w_l -y \frac{\partial}{\partial s} w_{yd} +\frac{x^2-y^2}{2} \frac{\partial}{\partial s} w_{yq}.
\end{align}
Next we calculate the derivatives of $w_{yd}$ and $w_{yq}$ with respect to $s$.
For each of these two functions there exist two analytical expressions, one for the double-sided~\cite{Bane2016} and another for the single-sided~\cite{Bane2016a} corrugated PS variant.
Conveniently, all four expressions have the form
\begin{align}
	w(s) = A \left[1 - \left(1 + \sqrt{\frac{s}{s_0}}\right) \exp\left(-\sqrt{\frac{s}{s_0}}\right)\right],
\end{align}
with $A$ and $s_0$ being functions of the corrugation geometry and the beam center of mass position inside the PS, and derivative
\begin{align}
	\frac{\partial w(s)}{\partial s} = \frac{A}{2s_0}\exp\left(-\sqrt{\frac{s}{s_0}}\right).
\end{align}
The energy change of an individual particle in the bunch is given by a convolution with the beam longitudinal charge density $\rho(s)$:
\begin{align}
	&\Delta E(s, x, y) = L \int_{-\infty}^s \mathrm{d}s' \rho(s') w^*_l(s-s', x, y) \nonumber
	\\
	&\equiv L \left(W_l(s) -y\ W_{ld}(s) + \frac{x^2-y^2}{2} W_{lq}(s)\right),
	\label{eq:echange}
\end{align}
where we denote $W_{ld}$ and $W_{lq}$ as the convolutions of the corresponding $\partial w_y/\partial s$ terms with $\rho(s)$, and where $L$ is the length of the structure.

Assuming Gaussian transverse beam distributions characterized by $\sigma_x$ and $\sigma_y$, we calculate the scaling of the induced energy spread with the beam sizes.
We obtain the slice mean energy loss $\Delta E$ and rms energy spread increase $\Delta \sigma_E$ by integrating over the transverse beam distribution, as indicated by the angle brackets:
\begin{align}
	\Delta E(s) &= \langle \Delta E(s,x,y) \rangle \nonumber
	\\
	&= L \left(W_l(s) + W_{lq}(s)\frac{\sigma_x^2 - \sigma_y^2}{2}\right)
	\\
	\Delta\sigma^2_E(s) &= \langle \Delta E^2(s,x,y) \rangle - \langle \Delta E(s,x,y) \rangle^2 \nonumber
	\\
	&= L^2\left(W_{ld}^2(s)\sigma_y^2 + W_{lq}^2(s)\frac{\sigma_x^4+\sigma_y^4}{2}\right).
	\label{eq:induced_espread}
\end{align}
For symmetry reasons, the slice mean energy change $\Delta E$ does not depend on $W_{ld}$: for every particle that receives an energy increase from the $y\ W_{ld}$ term in Eq.~(\ref{eq:echange}), there exists another particle with opposite $y$ coordinate that receives an energy decrease of the same magnitude.
The quadrupole wakefield correction leads to an additional slice energy change if the beam is not round.
Both the dipole and quadrupole wakefield corrections cause an energy spread increase.
The simulated $\Delta\sigma_E$ in Fig.~\ref{fig:sim_meas} (d) is almost entirely from the $W_{ld}$ correction term, with a negligible contribution from $W_{lq}$.

\section{Slice energy resolution for an infinitely long, flat beam}
\label{sec:effective_long}

We analytically solve the time-resolved energy spread measurement of an infinitely long beam with constant beam current, constant energy chirp $h=\mathrm{d}\delta/\mathrm{d}t$, linear streaking $\mu=\mathrm{d}y/\mathrm{d}t$, and linear dispersion $D=\mathrm{d}x/\mathrm{d}\delta$.
The 4D density function at the screen, assuming Gaussian distributions in the natural coordinates $x$, $y$, and $\delta=(E-E_0)/E_0$, and without normalization factors, is
\begin{align}
	\rho_{4D} = \exp\left( -\frac{(x-D\delta)^2}{2\sigma_x^2} -\frac{(y-\mu t)^2}{2\sigma_y^2} -\frac{(\delta-ht)^2}{2\sigma_\delta^2}\right).
	\label{eq:4d_density}
\end{align}
A screen records the projection $\rho_{2D}=\int \mathrm{d}t \int \mathrm{d}\delta\ \rho_{4d}$.
We evaluate the horizontal rms beam size $\sigma_{xm}$ along the streaked (vertical) dimension:
\begin{align}
	\sigma^2_{xm} &= \frac{\int \mathrm{d}x\ x^2 \rho_{2D}(x,y)}{\int \mathrm{d}x\ \rho_{2D}(x,y)} - \left(\frac{\int\mathrm{d}x\ x \rho_{2D}(x,y)}{\int \mathrm{d}x\ \rho_{2D}(x,y)}\right)^2 \nonumber
	\\
	&= \sigma_x^2 + \left(D\sigma_\delta \right)^2 + \left(\frac{Dh\sigma_y}{\mu}\right)^2,
	\label{eq:meas_beam_size}
\end{align}
and identify the measured energy spread $\sigma_{Em}$ as $E_0\sigma_{xm}/D$.
Attributing the difference between $\sigma_{Em}$ and the true energy spread $E_0\sigma_\delta$ to resolution effects, we obtain the slice energy resolution
\begin{align}
	r^*_E = \sqrt{\left(\frac{E_0\sigma_x}{D}\right)^2 + \left(\frac{\sigma_y}{\mu}\frac{\mathrm{d}E}{\mathrm{d}t}\right)^2},
\end{align}
equal to Eq.~(\ref{eq:energy_res3}) with $\Delta \sigma_E=0$ and $r_t=\sigma_y/|\mu|$.

\section{Comparison of tracking codes}
\label{sec:benchmark}

\begin{figure}
	\centering
	\includegraphics{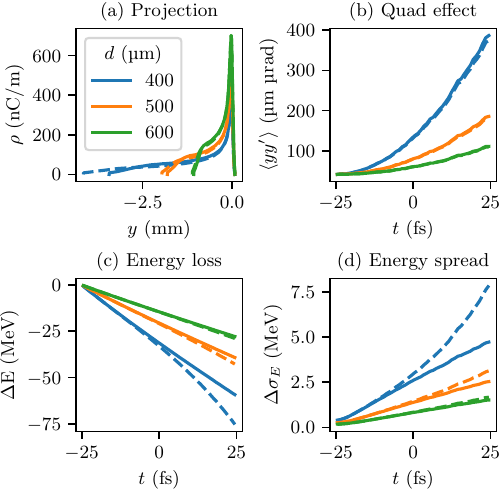}
	\caption{Comparison of the PWFM (solid) and OCELOT codes (dashed) in tracking a particle distribution from the PS to the screen.
(a): simulated streaked transverse distribution at the screen.
(b): evaluation of the time-resolved transverse quadrupole effect in streaking direction.
(c-d): slice energy loss and slice energy spread.
	}
	\label{fig:ocelot}
\end{figure}

For simulation results shown earlier in this paper we use the particle tracking functionality of the custom PWFM code~\cite{Dijkstal2022b}.
It implements the aforementioned analytical wakefield model~\cite{Bane2016}, and also the longitudinal wakefield corrections given in appendix~\ref{sec:long_wake}.
The code treats the beam as rigid inside the PS, and models the PS as a combination of a half-length drift, a point kick, and another half-length drift.
Therefore, each wakefield potential needs to be only calculated once, which is advantageous in terms of tracking speed.
However, once the wakefields become too strong, clearly the rigid beam approximation is no longer accurate.

Here we compare PWFM to the OCELOT code~\cite{Agapov2014}, which models a flat corrugated PS as many point kick sources interleaved by drift spaces, and thus resolves the dynamics of the beam inside the corrugated structure.
It implements the modal summation method for rectangular corrugated structures~\cite{Zagorodnov2016} and respects the Panofsky-Wenzel theorem on a fundamental level.
It contains also a third transverse wakefield component, which is relevant for beams that are tilted inside the structure~\cite{Qin2023}.

We generate a 6D particle distribution as described in Sec.~\ref{sec:sim_res}.
The initial beam has zero energy spread.
The step size for OCELOT is set to 0.1~m.
We use $d$ values of 400, 500, and 600~\textmu{}m to find the limit of PWFM's accuracy.
Figure~\ref{fig:ocelot} shows the analysis of the two tracking simulations.
We compare the simulated transverse particle distribution at the screen in (a).
The quadrupole effect increases the correlation between vertical positions and angles ($\langle yy'\rangle$).
We evaluate this quantity directly after the wakefield effects are applied for different time slices, and display the result in (b).
A comparison of the wakefield-induced energy loss and energy spread is shown in plots (c-d).
There is generally excellent agreement between the two codes for $d=600$~\textmu{}m and $d=500$~\textmu{}m.
We only see significant differences in the energy loss and energy spread at $d=400$~\textmu{}m, and emphasize that all experiments shown earlier in this paper were performed with $d$ larger than 500~\textmu{}m.


\bibliography{./main.bib}

\begin{thebibliography}{47}%
\makeatletter
\providecommand \@ifxundefined [1]{%
 \@ifx{#1\undefined}
}%
\providecommand \@ifnum [1]{%
 \ifnum #1\expandafter \@firstoftwo
 \else \expandafter \@secondoftwo
 \fi
}%
\providecommand \@ifx [1]{%
 \ifx #1\expandafter \@firstoftwo
 \else \expandafter \@secondoftwo
 \fi
}%
\providecommand \natexlab [1]{#1}%
\providecommand \enquote  [1]{``#1''}%
\providecommand \bibnamefont  [1]{#1}%
\providecommand \bibfnamefont [1]{#1}%
\providecommand \citenamefont [1]{#1}%
\providecommand \href@noop [0]{\@secondoftwo}%
\providecommand \href [0]{\begingroup \@sanitize@url \@href}%
\providecommand \@href[1]{\@@startlink{#1}\@@href}%
\providecommand \@@href[1]{\endgroup#1\@@endlink}%
\providecommand \@sanitize@url [0]{\catcode `\\12\catcode `\$12\catcode
  `\&12\catcode `\#12\catcode `\^12\catcode `\_12\catcode `\%12\relax}%
\providecommand \@@startlink[1]{}%
\providecommand \@@endlink[0]{}%
\providecommand \url  [0]{\begingroup\@sanitize@url \@url }%
\providecommand \@url [1]{\endgroup\@href {#1}{\urlprefix }}%
\providecommand \urlprefix  [0]{URL }%
\providecommand \Eprint [0]{\href }%
\providecommand \doibase [0]{https://doi.org/}%
\providecommand \selectlanguage [0]{\@gobble}%
\providecommand \bibinfo  [0]{\@secondoftwo}%
\providecommand \bibfield  [0]{\@secondoftwo}%
\providecommand \translation [1]{[#1]}%
\providecommand \BibitemOpen [0]{}%
\providecommand \bibitemStop [0]{}%
\providecommand \bibitemNoStop [0]{.\EOS\space}%
\providecommand \EOS [0]{\spacefactor3000\relax}%
\providecommand \BibitemShut  [1]{\csname bibitem#1\endcsname}%
\let\auto@bib@innerbib\@empty
\bibitem [{\citenamefont {Pellegrini}\ \emph {et~al.}(2016)\citenamefont
  {Pellegrini}, \citenamefont {Marinelli},\ and\ \citenamefont
  {Reiche}}]{Pellegrini2016}%
  \BibitemOpen
  \bibfield  {author} {\bibinfo {author} {\bibfnamefont {C.}~\bibnamefont
  {Pellegrini}}, \bibinfo {author} {\bibfnamefont {A.}~\bibnamefont
  {Marinelli}},\ and\ \bibinfo {author} {\bibfnamefont {S.}~\bibnamefont
  {Reiche}},\ }\bibfield  {title} {\bibinfo {title} {The physics of {x}-ray
  free-electron lasers},\ }\href {https://doi.org/10.1103/RevModPhys.88.015006}
  {\bibfield  {journal} {\bibinfo  {journal} {Rev. Mod. Phys.}\ }\textbf
  {\bibinfo {volume} {88}},\ \bibinfo {pages} {015006} (\bibinfo {year}
  {2016})}\BibitemShut {NoStop}%
\bibitem [{\citenamefont {Huang}\ \emph {et~al.}(2021)\citenamefont {Huang},
  \citenamefont {Deng}, \citenamefont {Liu}, \citenamefont {Wang},\ and\
  \citenamefont {Zhao}}]{Huang2021}%
  \BibitemOpen
  \bibfield  {author} {\bibinfo {author} {\bibfnamefont {N.}~\bibnamefont
  {Huang}}, \bibinfo {author} {\bibfnamefont {H.}~\bibnamefont {Deng}},
  \bibinfo {author} {\bibfnamefont {B.}~\bibnamefont {Liu}}, \bibinfo {author}
  {\bibfnamefont {D.}~\bibnamefont {Wang}},\ and\ \bibinfo {author}
  {\bibfnamefont {Z.}~\bibnamefont {Zhao}},\ }\bibfield  {title} {\bibinfo
  {title} {Features and futures of {X}-ray free-electron lasers},\ }\href
  {https://doi.org/10.1016/j.xinn.2021.100097} {\bibfield  {journal} {\bibinfo
  {journal} {Innovation}\ }\textbf {\bibinfo {volume} {2}},\ \bibinfo {pages}
  {100097} (\bibinfo {year} {2021})}\BibitemShut {NoStop}%
\bibitem [{\citenamefont {Tomin}\ \emph {et~al.}(2024)\citenamefont {Tomin},
  \citenamefont {Kaiser}, \citenamefont {Lockmann}, \citenamefont
  {Wohlenberg},\ and\ \citenamefont {Zagorodnov}}]{Tomin2024}%
  \BibitemOpen
  \bibfield  {author} {\bibinfo {author} {\bibfnamefont {S.}~\bibnamefont
  {Tomin}}, \bibinfo {author} {\bibfnamefont {J.}~\bibnamefont {Kaiser}},
  \bibinfo {author} {\bibfnamefont {N.~M.}\ \bibnamefont {Lockmann}}, \bibinfo
  {author} {\bibfnamefont {T.}~\bibnamefont {Wohlenberg}},\ and\ \bibinfo
  {author} {\bibfnamefont {I.}~\bibnamefont {Zagorodnov}},\ }\bibfield  {title}
  {\bibinfo {title} {Undulator linear taper control at the {E}uropean {X}-ray
  {F}ree-{E}lectron {L}aser facility},\ }\href
  {https://doi.org/10.1103/PhysRevAccelBeams.27.042801} {\bibfield  {journal}
  {\bibinfo  {journal} {Phys. Rev. Accel. Beams}\ }\textbf {\bibinfo {volume}
  {27}},\ \bibinfo {pages} {042801} (\bibinfo {year} {2024})}\BibitemShut
  {NoStop}%
\bibitem [{\citenamefont {Fadini}\ \emph {et~al.}(2020)\citenamefont {Fadini},
  \citenamefont {Reiche}, \citenamefont {Nass},\ and\ \citenamefont {van
  Thor}}]{Fadini2020}%
  \BibitemOpen
  \bibfield  {author} {\bibinfo {author} {\bibfnamefont {A.}~\bibnamefont
  {Fadini}}, \bibinfo {author} {\bibfnamefont {S.}~\bibnamefont {Reiche}},
  \bibinfo {author} {\bibfnamefont {K.}~\bibnamefont {Nass}},\ and\ \bibinfo
  {author} {\bibfnamefont {J.~J.}\ \bibnamefont {van Thor}},\ }\bibfield
  {title} {\bibinfo {title} {Applications and limits of time-to-energy mapping
  of protein crystal diffraction using energy-chirped polychromatic {XFEL}
  pulses},\ }\bibfield  {journal} {\bibinfo  {journal} {Appl. Sci.}\ }\textbf
  {\bibinfo {volume} {10}},\ \href {https://doi.org/10.3390/app10072599}
  {10.3390/app10072599} (\bibinfo {year} {2020})\BibitemShut {NoStop}%
\bibitem [{\citenamefont {Nass}\ \emph {et~al.}(2021)\citenamefont {Nass},
  \citenamefont {Bacellar}, \citenamefont {Cirelli}, \citenamefont
  {Dworkowski}, \citenamefont {Gevorkov}, \citenamefont {James}, \citenamefont
  {Johnson}, \citenamefont {Kekilli}, \citenamefont {Knopp}, \citenamefont
  {Martiel} \emph {et~al.}}]{Nass2021}%
  \BibitemOpen
  \bibfield  {author} {\bibinfo {author} {\bibfnamefont {K.}~\bibnamefont
  {Nass}}, \bibinfo {author} {\bibfnamefont {C.}~\bibnamefont {Bacellar}},
  \bibinfo {author} {\bibfnamefont {C.}~\bibnamefont {Cirelli}}, \bibinfo
  {author} {\bibfnamefont {F.}~\bibnamefont {Dworkowski}}, \bibinfo {author}
  {\bibfnamefont {Y.}~\bibnamefont {Gevorkov}}, \bibinfo {author}
  {\bibfnamefont {D.}~\bibnamefont {James}}, \bibinfo {author} {\bibfnamefont
  {P.~J.~M.}\ \bibnamefont {Johnson}}, \bibinfo {author} {\bibfnamefont
  {D.}~\bibnamefont {Kekilli}}, \bibinfo {author} {\bibfnamefont
  {G.}~\bibnamefont {Knopp}}, \bibinfo {author} {\bibfnamefont
  {I.}~\bibnamefont {Martiel}}, \emph {et~al.},\ }\bibfield  {title} {\bibinfo
  {title} {Pink-beam serial femtosecond crystallography for accurate
  structure-factor determination at an {X}-ray free-electron laser},\ }\href
  {https://doi.org/10.1107/S2052252521008046} {\bibfield  {journal} {\bibinfo
  {journal} {IUCrJ}\ }\textbf {\bibinfo {volume} {8}},\ \bibinfo {pages} {905}
  (\bibinfo {year} {2021})}\BibitemShut {NoStop}%
\bibitem [{\citenamefont {Kondratenko}\ and\ \citenamefont
  {Saldin}(1980)}]{Kondratenko1980}%
  \BibitemOpen
  \bibfield  {author} {\bibinfo {author} {\bibfnamefont {A.~M.}\ \bibnamefont
  {Kondratenko}}\ and\ \bibinfo {author} {\bibfnamefont {E.~L.}\ \bibnamefont
  {Saldin}},\ }\bibfield  {title} {\bibinfo {title} {Generation of coherent
  radiation by a relativistic electron beam in an ondulator},\ }\href
  {https://cds.cern.ch/record/1107977} {\bibfield  {journal} {\bibinfo
  {journal} {Particle Accelerators}\ }\textbf {\bibinfo {volume} {10}},\
  \bibinfo {pages} {207} (\bibinfo {year} {1980})}\BibitemShut {NoStop}%
\bibitem [{\citenamefont {Bonifacio}\ \emph {et~al.}(1984)\citenamefont
  {Bonifacio}, \citenamefont {Pellegrini},\ and\ \citenamefont
  {Narducci}}]{Bonifacio1984}%
  \BibitemOpen
  \bibfield  {author} {\bibinfo {author} {\bibfnamefont {R.}~\bibnamefont
  {Bonifacio}}, \bibinfo {author} {\bibfnamefont {C.}~\bibnamefont
  {Pellegrini}},\ and\ \bibinfo {author} {\bibfnamefont {L.~M.}\ \bibnamefont
  {Narducci}},\ }\bibfield  {title} {\bibinfo {title} {Collective instabilities
  and high-gain regime in a free electron laser},\ }\href
  {https://doi.org/10.1016/0030-4018(84)90105-6} {\bibfield  {journal}
  {\bibinfo  {journal} {Opt. Commun.}\ }\textbf {\bibinfo {volume} {50}},\
  \bibinfo {pages} {373 } (\bibinfo {year} {1984})}\BibitemShut {NoStop}%
\bibitem [{\citenamefont {Saldin}\ \emph {et~al.}(2004)\citenamefont {Saldin},
  \citenamefont {Schneidmiller},\ and\ \citenamefont {Yurkov}}]{Saldin2004}%
  \BibitemOpen
  \bibfield  {author} {\bibinfo {author} {\bibfnamefont {E.~L.}\ \bibnamefont
  {Saldin}}, \bibinfo {author} {\bibfnamefont {E.~A.}\ \bibnamefont
  {Schneidmiller}},\ and\ \bibinfo {author} {\bibfnamefont {M.~V.}\
  \bibnamefont {Yurkov}},\ }\bibfield  {title} {\bibinfo {title} {Longitudinal
  space charge-driven microbunching instability in the {TESLA Test Facility}
  linac},\ }\href {https://doi.org/10.1016/j.nima.2004.04.067} {\bibfield
  {journal} {\bibinfo  {journal} {Nucl. Instrum. Methods Phys. Res., Sect. A}\
  }\textbf {\bibinfo {volume} {528}},\ \bibinfo {pages} {355} (\bibinfo {year}
  {2004})}\BibitemShut {NoStop}%
\bibitem [{\citenamefont {Ding}\ \emph {et~al.}(2011)\citenamefont {Ding},
  \citenamefont {Behrens}, \citenamefont {Emma}, \citenamefont {Frisch},
  \citenamefont {Huang}, \citenamefont {Loos}, \citenamefont {Krejcik},\ and\
  \citenamefont {Wang}}]{Ding2011}%
  \BibitemOpen
  \bibfield  {author} {\bibinfo {author} {\bibfnamefont {Y.}~\bibnamefont
  {Ding}}, \bibinfo {author} {\bibfnamefont {C.}~\bibnamefont {Behrens}},
  \bibinfo {author} {\bibfnamefont {P.}~\bibnamefont {Emma}}, \bibinfo {author}
  {\bibfnamefont {J.}~\bibnamefont {Frisch}}, \bibinfo {author} {\bibfnamefont
  {Z.}~\bibnamefont {Huang}}, \bibinfo {author} {\bibfnamefont
  {H.}~\bibnamefont {Loos}}, \bibinfo {author} {\bibfnamefont {P.}~\bibnamefont
  {Krejcik}},\ and\ \bibinfo {author} {\bibfnamefont {M.-H.}\ \bibnamefont
  {Wang}},\ }\bibfield  {title} {\bibinfo {title} {Femtosecond {X}-ray pulse
  temporal characterization in free-electron lasers using a transverse
  deflector},\ }\href {https://doi.org/10.1103/physrevstab.14.120701}
  {\bibfield  {journal} {\bibinfo  {journal} {Phys. Rev. Spec. Top. Accel.
  Beams}\ }\textbf {\bibinfo {volume} {14}},\ \bibinfo {pages} {120701}
  (\bibinfo {year} {2011})}\BibitemShut {NoStop}%
\bibitem [{\citenamefont {Behrens}\ \emph {et~al.}(2014)\citenamefont
  {Behrens}, \citenamefont {Decker}, \citenamefont {Ding}, \citenamefont
  {Dolgashev}, \citenamefont {Frisch}, \citenamefont {Huang}, \citenamefont
  {Krejcik}, \citenamefont {Loos}, \citenamefont {Lutman}, \citenamefont
  {Maxwell} \emph {et~al.}}]{Behrens2014}%
  \BibitemOpen
  \bibfield  {author} {\bibinfo {author} {\bibfnamefont {C.}~\bibnamefont
  {Behrens}}, \bibinfo {author} {\bibfnamefont {F.-J.}\ \bibnamefont {Decker}},
  \bibinfo {author} {\bibfnamefont {Y.}~\bibnamefont {Ding}}, \bibinfo {author}
  {\bibfnamefont {V.~A.}\ \bibnamefont {Dolgashev}}, \bibinfo {author}
  {\bibfnamefont {J.}~\bibnamefont {Frisch}}, \bibinfo {author} {\bibfnamefont
  {Z.}~\bibnamefont {Huang}}, \bibinfo {author} {\bibfnamefont
  {P.}~\bibnamefont {Krejcik}}, \bibinfo {author} {\bibfnamefont
  {H.}~\bibnamefont {Loos}}, \bibinfo {author} {\bibfnamefont {A.}~\bibnamefont
  {Lutman}}, \bibinfo {author} {\bibfnamefont {T.~J.}\ \bibnamefont {Maxwell}},
  \emph {et~al.},\ }\bibfield  {title} {\bibinfo {title} {Few-femtosecond
  time-resolved measurements of {X}-ray free-electron lasers},\ }\href
  {https://doi.org/10.1038/ncomms4762} {\bibfield  {journal} {\bibinfo
  {journal} {Nat. Commun.}\ }\textbf {\bibinfo {volume} {5}},\ \bibinfo {pages}
  {3762} (\bibinfo {year} {2014})}\BibitemShut {NoStop}%
\bibitem [{\citenamefont {Dijkstal}\ \emph {et~al.}(2022)\citenamefont
  {Dijkstal}, \citenamefont {Malyzhenkov}, \citenamefont {Craievich},
  \citenamefont {Ferrari}, \citenamefont {Ganter}, \citenamefont {Reiche},
  \citenamefont {Schietinger}, \citenamefont {Jurani\'{c}},\ and\ \citenamefont
  {Prat}}]{Dijkstal2022}%
  \BibitemOpen
  \bibfield  {author} {\bibinfo {author} {\bibfnamefont {P.}~\bibnamefont
  {Dijkstal}}, \bibinfo {author} {\bibfnamefont {A.}~\bibnamefont
  {Malyzhenkov}}, \bibinfo {author} {\bibfnamefont {P.}~\bibnamefont
  {Craievich}}, \bibinfo {author} {\bibfnamefont {E.}~\bibnamefont {Ferrari}},
  \bibinfo {author} {\bibfnamefont {R.}~\bibnamefont {Ganter}}, \bibinfo
  {author} {\bibfnamefont {S.}~\bibnamefont {Reiche}}, \bibinfo {author}
  {\bibfnamefont {T.}~\bibnamefont {Schietinger}}, \bibinfo {author}
  {\bibfnamefont {P.}~\bibnamefont {Jurani\'{c}}},\ and\ \bibinfo {author}
  {\bibfnamefont {E.}~\bibnamefont {Prat}},\ }\bibfield  {title} {\bibinfo
  {title} {Self-synchronized and cost-effective time-resolved measurements at
  {X}-ray free-electron lasers with femtosecond resolution},\ }\href
  {https://doi.org/10.1103/physrevresearch.4.013017} {\bibfield  {journal}
  {\bibinfo  {journal} {Phys. Rev. Res.}\ }\textbf {\bibinfo {volume} {4}},\
  \bibinfo {pages} {013017} (\bibinfo {year} {2022})}\BibitemShut {NoStop}%
\bibitem [{\citenamefont {Chapman}\ \emph {et~al.}(2014)\citenamefont
  {Chapman}, \citenamefont {Caleman},\ and\ \citenamefont
  {Timneanu}}]{Chapman2014}%
  \BibitemOpen
  \bibfield  {author} {\bibinfo {author} {\bibfnamefont {H.~N.}\ \bibnamefont
  {Chapman}}, \bibinfo {author} {\bibfnamefont {C.}~\bibnamefont {Caleman}},\
  and\ \bibinfo {author} {\bibfnamefont {N.}~\bibnamefont {Timneanu}},\
  }\bibfield  {title} {\bibinfo {title} {Diffraction before destruction},\
  }\href {https://doi.org/10.1098/rstb.2013.0313} {\bibfield  {journal}
  {\bibinfo  {journal} {Philos. Trans. R. Soc. London, Ser. B}\ }\textbf
  {\bibinfo {volume} {369}},\ \bibinfo {pages} {20130313} (\bibinfo {year}
  {2014})}\BibitemShut {NoStop}%
\bibitem [{\citenamefont {Young}\ \emph {et~al.}(2018)\citenamefont {Young},
  \citenamefont {Ueda}, \citenamefont {Gühr}, \citenamefont {Bucksbaum},
  \citenamefont {Simon}, \citenamefont {Mukamel}, \citenamefont {Rohringer},
  \citenamefont {Prince}, \citenamefont {Masciovecchio}, \citenamefont {Meyer}
  \emph {et~al.}}]{Young2018}%
  \BibitemOpen
  \bibfield  {author} {\bibinfo {author} {\bibfnamefont {L.}~\bibnamefont
  {Young}}, \bibinfo {author} {\bibfnamefont {K.}~\bibnamefont {Ueda}},
  \bibinfo {author} {\bibfnamefont {M.}~\bibnamefont {Gühr}}, \bibinfo
  {author} {\bibfnamefont {P.~H.}\ \bibnamefont {Bucksbaum}}, \bibinfo {author}
  {\bibfnamefont {M.}~\bibnamefont {Simon}}, \bibinfo {author} {\bibfnamefont
  {S.}~\bibnamefont {Mukamel}}, \bibinfo {author} {\bibfnamefont
  {N.}~\bibnamefont {Rohringer}}, \bibinfo {author} {\bibfnamefont {K.~C.}\
  \bibnamefont {Prince}}, \bibinfo {author} {\bibfnamefont {C.}~\bibnamefont
  {Masciovecchio}}, \bibinfo {author} {\bibfnamefont {M.}~\bibnamefont
  {Meyer}}, \emph {et~al.},\ }\bibfield  {title} {\bibinfo {title} {Roadmap of
  ultrafast {X}-ray atomic and molecular physics},\ }\href
  {https://doi.org/10.1088/1361-6455/aa9735} {\bibfield  {journal} {\bibinfo
  {journal} {J. Phys. B: At. Mol. Opt. Phys.}\ }\textbf {\bibinfo {volume}
  {51}},\ \bibinfo {pages} {032003} (\bibinfo {year} {2018})}\BibitemShut
  {NoStop}%
\bibitem [{\citenamefont {Chapman}(2019)}]{Chapman2019}%
  \BibitemOpen
  \bibfield  {author} {\bibinfo {author} {\bibfnamefont {H.~N.}\ \bibnamefont
  {Chapman}},\ }\bibfield  {title} {\bibinfo {title} {X-ray free-electron
  lasers for the structure and dynamics of macromolecules},\ }\href
  {https://doi.org/10.1146/annurev-biochem-013118-110744} {\bibfield  {journal}
  {\bibinfo  {journal} {Annu. Rev. Biochem.}\ }\textbf {\bibinfo {volume}
  {88}},\ \bibinfo {pages} {35} (\bibinfo {year} {2019})}\BibitemShut {NoStop}%
\bibitem [{\citenamefont {Emma}\ \emph {et~al.}(2000)\citenamefont {Emma},
  \citenamefont {Frisch},\ and\ \citenamefont {Krejcik}}]{Emma2000}%
  \BibitemOpen
  \bibfield  {author} {\bibinfo {author} {\bibfnamefont {P.}~\bibnamefont
  {Emma}}, \bibinfo {author} {\bibfnamefont {J.}~\bibnamefont {Frisch}},\ and\
  \bibinfo {author} {\bibfnamefont {P.}~\bibnamefont {Krejcik}},\ }\href
  {https://www-ssrl.slac.stanford.edu/lcls/technotes/lcls-tn-00-12.pdf} {\emph
  {\bibinfo {title} {A transverse rf deflecting structure for bunch length and
  phase space diagnostics}}},\ \bibinfo {type} {LCLS Report}\ \bibinfo {number}
  {LCLS-TN-00-12}\ (\bibinfo  {institution} {Stanford Linear Accelerator
  Center},\ \bibinfo {year} {2000})\BibitemShut {NoStop}%
\bibitem [{\citenamefont {Floettmann}\ and\ \citenamefont
  {Paramonov}(2014)}]{Floettmann2014}%
  \BibitemOpen
  \bibfield  {author} {\bibinfo {author} {\bibfnamefont {K.}~\bibnamefont
  {Floettmann}}\ and\ \bibinfo {author} {\bibfnamefont {V.~V.}\ \bibnamefont
  {Paramonov}},\ }\bibfield  {title} {\bibinfo {title} {Beam dynamics in
  transverse deflecting rf structures},\ }\href
  {https://doi.org/10.1103/physrevstab.17.024001} {\bibfield  {journal}
  {\bibinfo  {journal} {Phys. Rev. Spec. Top. Accel. Beams}\ }\textbf {\bibinfo
  {volume} {17}},\ \bibinfo {pages} {024001} (\bibinfo {year}
  {2014})}\BibitemShut {NoStop}%
\bibitem [{\citenamefont {Wiebers}\ \emph {et~al.}(2013)\citenamefont
  {Wiebers}, \citenamefont {Holz}, \citenamefont {Kube}, \citenamefont
  {N\"{o}lle}, \citenamefont {Priebe},\ and\ \citenamefont
  {Schr\"{o}der}}]{Wiebers2013}%
  \BibitemOpen
  \bibfield  {author} {\bibinfo {author} {\bibfnamefont {C.}~\bibnamefont
  {Wiebers}}, \bibinfo {author} {\bibfnamefont {M.}~\bibnamefont {Holz}},
  \bibinfo {author} {\bibfnamefont {G.}~\bibnamefont {Kube}}, \bibinfo {author}
  {\bibfnamefont {D.}~\bibnamefont {N\"{o}lle}}, \bibinfo {author}
  {\bibfnamefont {G.}~\bibnamefont {Priebe}},\ and\ \bibinfo {author}
  {\bibfnamefont {H.-C.}\ \bibnamefont {Schr\"{o}der}},\ }\bibfield  {title}
  {\bibinfo {title} {Scintillating screen monitors for transverse electron beam
  profile diagnostics at the {European XFEL}},\ }in\ \href
  {http://accelconf.web.cern.ch/IBIC2013/papers/wepf03.pdf} {\emph {\bibinfo
  {booktitle} {Proc. IBIC 2013, Oxford, United Kingdom}}},\ \bibinfo {editor}
  {edited by\ \bibinfo {editor} {\bibfnamefont {I.}~\bibnamefont {Martin}}\
  and\ \bibinfo {editor} {\bibfnamefont {G.}~\bibnamefont {Rehm}}}\ (\bibinfo
  {publisher} {JACoW},\ \bibinfo {address} {Geneva},\ \bibinfo {year} {2013})\
  \bibinfo {note} {{WEPF03}}\BibitemShut {NoStop}%
\bibitem [{\citenamefont {Bettoni}\ \emph {et~al.}(2016)\citenamefont
  {Bettoni}, \citenamefont {Craievich}, \citenamefont {Lutman},\ and\
  \citenamefont {Pedrozzi}}]{Bettoni2016}%
  \BibitemOpen
  \bibfield  {author} {\bibinfo {author} {\bibfnamefont {S.}~\bibnamefont
  {Bettoni}}, \bibinfo {author} {\bibfnamefont {P.}~\bibnamefont {Craievich}},
  \bibinfo {author} {\bibfnamefont {A.~A.}\ \bibnamefont {Lutman}},\ and\
  \bibinfo {author} {\bibfnamefont {M.}~\bibnamefont {Pedrozzi}},\ }\bibfield
  {title} {\bibinfo {title} {Temporal profile measurements of relativistic
  electron bunch based on wakefield generation},\ }\href
  {https://doi.org/10.1103/physrevaccelbeams.19.021304} {\bibfield  {journal}
  {\bibinfo  {journal} {Phys. Rev. Accel. Beams}\ }\textbf {\bibinfo {volume}
  {19}},\ \bibinfo {pages} {021304} (\bibinfo {year} {2016})}\BibitemShut
  {NoStop}%
\bibitem [{\citenamefont {Seok}\ \emph {et~al.}(2018)\citenamefont {Seok},
  \citenamefont {Chung}, \citenamefont {Kang}, \citenamefont {Min},\ and\
  \citenamefont {Na}}]{Seok2018}%
  \BibitemOpen
  \bibfield  {author} {\bibinfo {author} {\bibfnamefont {J.}~\bibnamefont
  {Seok}}, \bibinfo {author} {\bibfnamefont {M.}~\bibnamefont {Chung}},
  \bibinfo {author} {\bibfnamefont {H.-S.}\ \bibnamefont {Kang}}, \bibinfo
  {author} {\bibfnamefont {C.-K.}\ \bibnamefont {Min}},\ and\ \bibinfo {author}
  {\bibfnamefont {D.}~\bibnamefont {Na}},\ }\bibfield  {title} {\bibinfo
  {title} {Use of a corrugated beam pipe as a passive deflector for bunch
  length measurements},\ }\href
  {https://doi.org/10.1103/physrevaccelbeams.21.022801} {\bibfield  {journal}
  {\bibinfo  {journal} {Phys. Rev. Accel. Beams}\ }\textbf {\bibinfo {volume}
  {21}},\ \bibinfo {pages} {022801} (\bibinfo {year} {2018})}\BibitemShut
  {NoStop}%
\bibitem [{\citenamefont {Tomin}\ \emph {et~al.}(2022)\citenamefont {Tomin},
  \citenamefont {Decking}, \citenamefont {Golubeva}, \citenamefont
  {Novokshonov}, \citenamefont {Wohlenberg},\ and\ \citenamefont
  {Zagorodnov}}]{Tomin2022}%
  \BibitemOpen
  \bibfield  {author} {\bibinfo {author} {\bibfnamefont {S.}~\bibnamefont
  {Tomin}}, \bibinfo {author} {\bibfnamefont {W.}~\bibnamefont {Decking}},
  \bibinfo {author} {\bibfnamefont {N.}~\bibnamefont {Golubeva}}, \bibinfo
  {author} {\bibfnamefont {A.}~\bibnamefont {Novokshonov}}, \bibinfo {author}
  {\bibfnamefont {T.}~\bibnamefont {Wohlenberg}},\ and\ \bibinfo {author}
  {\bibfnamefont {I.}~\bibnamefont {Zagorodnov}},\ }\bibfield  {title}
  {\bibinfo {title} {Longitudinal phase space diagnostics with corrugated
  structure at the {European XFEL}},\ }in\ \href
  {https://doi.org/10.18429/JACOW-IPAC2022-MOPOPT020} {\emph {\bibinfo
  {booktitle} {Proc. IPAC 2022, Bangkok, Thailand}}},\ \bibinfo {editor}
  {edited by\ \bibinfo {editor} {\bibfnamefont {F.}~\bibnamefont {Zimmermann}},
  \bibinfo {editor} {\bibfnamefont {H.}~\bibnamefont {Tanaka}}, \bibinfo
  {editor} {\bibfnamefont {P.}~\bibnamefont {Sudmuang}}, \bibinfo {editor}
  {\bibfnamefont {P.}~\bibnamefont {Klysubun}}, \bibinfo {editor}
  {\bibfnamefont {P.}~\bibnamefont {Sunwong}}, \bibinfo {editor} {\bibfnamefont
  {T.}~\bibnamefont {Chanwattana}}, \bibinfo {editor} {\bibfnamefont
  {C.}~\bibnamefont {Petit-Jean-Genaz}},\ and\ \bibinfo {editor} {\bibfnamefont
  {V.~R.~W.}\ \bibnamefont {Schaa}}}\ (\bibinfo  {publisher} {JACoW},\ \bibinfo
  {address} {Geneva},\ \bibinfo {year} {2022})\ \bibinfo {note}
  {{MOPOPT020}}\BibitemShut {NoStop}%
\bibitem [{\citenamefont {Prat}\ \emph
  {et~al.}(2020{\natexlab{a}})\citenamefont {Prat}, \citenamefont {Abela},
  \citenamefont {Aiba}, \citenamefont {Alarcon}, \citenamefont {Alex},
  \citenamefont {Arbelo}, \citenamefont {Arrell}, \citenamefont {Arsov},
  \citenamefont {Bacellar}, \citenamefont {Beard} \emph {et~al.}}]{Prat2020b}%
  \BibitemOpen
  \bibfield  {author} {\bibinfo {author} {\bibfnamefont {E.}~\bibnamefont
  {Prat}}, \bibinfo {author} {\bibfnamefont {R.}~\bibnamefont {Abela}},
  \bibinfo {author} {\bibfnamefont {M.}~\bibnamefont {Aiba}}, \bibinfo {author}
  {\bibfnamefont {A.}~\bibnamefont {Alarcon}}, \bibinfo {author} {\bibfnamefont
  {J.}~\bibnamefont {Alex}}, \bibinfo {author} {\bibfnamefont {Y.}~\bibnamefont
  {Arbelo}}, \bibinfo {author} {\bibfnamefont {C.}~\bibnamefont {Arrell}},
  \bibinfo {author} {\bibfnamefont {V.}~\bibnamefont {Arsov}}, \bibinfo
  {author} {\bibfnamefont {C.}~\bibnamefont {Bacellar}}, \bibinfo {author}
  {\bibfnamefont {C.}~\bibnamefont {Beard}}, \emph {et~al.},\ }\bibfield
  {title} {\bibinfo {title} {A compact and cost-effective hard {X}-ray
  free-electron laser driven by a high-brightness and low-energy electron
  beam},\ }\href {https://doi.org/10.1038/s41566-020-00712-8} {\bibfield
  {journal} {\bibinfo  {journal} {Nat. Photonics}\ }\textbf {\bibinfo {volume}
  {14}},\ \bibinfo {pages} {748} (\bibinfo {year}
  {2020}{\natexlab{a}})}\BibitemShut {NoStop}%
\bibitem [{\citenamefont {Decking}\ \emph {et~al.}(2020)\citenamefont
  {Decking}, \citenamefont {Abeghyan}, \citenamefont {Abramian}, \citenamefont
  {Abramsky}, \citenamefont {Aguirre}, \citenamefont {Albrecht}, \citenamefont
  {Alou}, \citenamefont {Altarelli}, \citenamefont {Altmann}, \citenamefont
  {Amyan} \emph {et~al.}}]{Decking2020}%
  \BibitemOpen
  \bibfield  {author} {\bibinfo {author} {\bibfnamefont {W.}~\bibnamefont
  {Decking}}, \bibinfo {author} {\bibfnamefont {S.}~\bibnamefont {Abeghyan}},
  \bibinfo {author} {\bibfnamefont {P.}~\bibnamefont {Abramian}}, \bibinfo
  {author} {\bibfnamefont {A.}~\bibnamefont {Abramsky}}, \bibinfo {author}
  {\bibfnamefont {A.}~\bibnamefont {Aguirre}}, \bibinfo {author} {\bibfnamefont
  {C.}~\bibnamefont {Albrecht}}, \bibinfo {author} {\bibfnamefont
  {P.}~\bibnamefont {Alou}}, \bibinfo {author} {\bibfnamefont {M.}~\bibnamefont
  {Altarelli}}, \bibinfo {author} {\bibfnamefont {P.}~\bibnamefont {Altmann}},
  \bibinfo {author} {\bibfnamefont {K.}~\bibnamefont {Amyan}}, \emph {et~al.},\
  }\bibfield  {title} {\bibinfo {title} {A {MHz}-repetition-rate hard {X}-ray
  free-electron laser driven by a superconducting linear accelerator},\ }\href
  {https://doi.org/10.1038/s41566-020-0607-z} {\bibfield  {journal} {\bibinfo
  {journal} {Nat. Photonics}\ }\textbf {\bibinfo {volume} {14}},\ \bibinfo
  {pages} {391} (\bibinfo {year} {2020})}\BibitemShut {NoStop}%
\bibitem [{\citenamefont {Novokhatski}(2015)}]{Novokhatski2015}%
  \BibitemOpen
  \bibfield  {author} {\bibinfo {author} {\bibfnamefont {A.}~\bibnamefont
  {Novokhatski}},\ }\bibfield  {title} {\bibinfo {title} {Wakefield potentials
  of corrugated structures},\ }\href
  {https://doi.org/10.1103/physrevstab.18.104402} {\bibfield  {journal}
  {\bibinfo  {journal} {Phys. Rev. Spec. Top. Accel. Beams}\ }\textbf {\bibinfo
  {volume} {18}},\ \bibinfo {pages} {104402} (\bibinfo {year}
  {2015})}\BibitemShut {NoStop}%
\bibitem [{\citenamefont {Huang}\ \emph {et~al.}(2017)\citenamefont {Huang},
  \citenamefont {Ding}, \citenamefont {Feng}, \citenamefont {Hemsing},
  \citenamefont {Huang}, \citenamefont {Krzywinski}, \citenamefont {Lutman},
  \citenamefont {Marinelli}, \citenamefont {Maxwell},\ and\ \citenamefont
  {Zhu}}]{Huang2017}%
  \BibitemOpen
  \bibfield  {author} {\bibinfo {author} {\bibfnamefont {S.}~\bibnamefont
  {Huang}}, \bibinfo {author} {\bibfnamefont {Y.}~\bibnamefont {Ding}},
  \bibinfo {author} {\bibfnamefont {Y.}~\bibnamefont {Feng}}, \bibinfo {author}
  {\bibfnamefont {E.}~\bibnamefont {Hemsing}}, \bibinfo {author} {\bibfnamefont
  {Z.}~\bibnamefont {Huang}}, \bibinfo {author} {\bibfnamefont
  {J.}~\bibnamefont {Krzywinski}}, \bibinfo {author} {\bibfnamefont {A.~A.}\
  \bibnamefont {Lutman}}, \bibinfo {author} {\bibfnamefont {A.}~\bibnamefont
  {Marinelli}}, \bibinfo {author} {\bibfnamefont {T.~J.}\ \bibnamefont
  {Maxwell}},\ and\ \bibinfo {author} {\bibfnamefont {D.}~\bibnamefont {Zhu}},\
  }\bibfield  {title} {\bibinfo {title} {Generating single-spike hard x-ray
  pulses with nonlinear bunch compression in free-electron lasers},\ }\href
  {https://doi.org/10.1103/PhysRevLett.119.154801} {\bibfield  {journal}
  {\bibinfo  {journal} {Phys. Rev. Lett.}\ }\textbf {\bibinfo {volume} {119}},\
  \bibinfo {pages} {154801} (\bibinfo {year} {2017})}\BibitemShut {NoStop}%
\bibitem [{\citenamefont {Trebushinin}\ \emph {et~al.}(2023)\citenamefont
  {Trebushinin}, \citenamefont {Geloni}, \citenamefont {Serkez}, \citenamefont
  {Mercurio}, \citenamefont {Gerasimova}, \citenamefont {Maltezopoulos},
  \citenamefont {Guetg},\ and\ \citenamefont
  {Schneidmiller}}]{Trebushinin2023}%
  \BibitemOpen
  \bibfield  {author} {\bibinfo {author} {\bibfnamefont {A.}~\bibnamefont
  {Trebushinin}}, \bibinfo {author} {\bibfnamefont {G.}~\bibnamefont {Geloni}},
  \bibinfo {author} {\bibfnamefont {S.}~\bibnamefont {Serkez}}, \bibinfo
  {author} {\bibfnamefont {G.}~\bibnamefont {Mercurio}}, \bibinfo {author}
  {\bibfnamefont {N.}~\bibnamefont {Gerasimova}}, \bibinfo {author}
  {\bibfnamefont {T.}~\bibnamefont {Maltezopoulos}}, \bibinfo {author}
  {\bibfnamefont {M.}~\bibnamefont {Guetg}},\ and\ \bibinfo {author}
  {\bibfnamefont {E.}~\bibnamefont {Schneidmiller}},\ }\bibfield  {title}
  {\bibinfo {title} {Experimental demonstration of attoseconds-at-harmonics at
  the {SASE3} undulator of the {European XFEL}},\ }\bibfield  {journal}
  {\bibinfo  {journal} {Photonics}\ }\textbf {\bibinfo {volume} {10}},\ \href
  {https://doi.org/10.3390/photonics10020131} {10.3390/photonics10020131}
  (\bibinfo {year} {2023})\BibitemShut {NoStop}%
\bibitem [{\citenamefont {Penco}\ \emph {et~al.}(2017)\citenamefont {Penco},
  \citenamefont {Allaria}, \citenamefont {De~Ninno}, \citenamefont {Ferrari},
  \citenamefont {Giannessi}, \citenamefont {Roussel},\ and\ \citenamefont
  {Spampinati}}]{Penco2017}%
  \BibitemOpen
  \bibfield  {author} {\bibinfo {author} {\bibfnamefont {G.}~\bibnamefont
  {Penco}}, \bibinfo {author} {\bibfnamefont {E.}~\bibnamefont {Allaria}},
  \bibinfo {author} {\bibfnamefont {G.}~\bibnamefont {De~Ninno}}, \bibinfo
  {author} {\bibfnamefont {E.}~\bibnamefont {Ferrari}}, \bibinfo {author}
  {\bibfnamefont {L.}~\bibnamefont {Giannessi}}, \bibinfo {author}
  {\bibfnamefont {E.}~\bibnamefont {Roussel}},\ and\ \bibinfo {author}
  {\bibfnamefont {S.}~\bibnamefont {Spampinati}},\ }\bibfield  {title}
  {\bibinfo {title} {Optical klystron enhancement to self ampliﬁed
  spontaneous emission at {FERMI}},\ }\bibfield  {journal} {\bibinfo  {journal}
  {Photonics}\ }\textbf {\bibinfo {volume} {4}},\ \href
  {https://doi.org/10.3390/photonics4010015} {10.3390/photonics4010015}
  (\bibinfo {year} {2017})\BibitemShut {NoStop}%
\bibitem [{\citenamefont {Allaria}\ \emph {et~al.}(2012)\citenamefont
  {Allaria}, \citenamefont {Castronovo}, \citenamefont {Cinquegrana},
  \citenamefont {Craievich}, \citenamefont {Dal~Forno}, \citenamefont
  {Danailov}, \citenamefont {D{\textquotesingle}Auria}, \citenamefont
  {Demidovich}, \citenamefont {De~Ninno}, \citenamefont {Di~Mitri} \emph
  {et~al.}}]{Allaria2012}%
  \BibitemOpen
  \bibfield  {author} {\bibinfo {author} {\bibfnamefont {E.}~\bibnamefont
  {Allaria}}, \bibinfo {author} {\bibfnamefont {D.}~\bibnamefont {Castronovo}},
  \bibinfo {author} {\bibfnamefont {P.}~\bibnamefont {Cinquegrana}}, \bibinfo
  {author} {\bibfnamefont {P.}~\bibnamefont {Craievich}}, \bibinfo {author}
  {\bibfnamefont {M.}~\bibnamefont {Dal~Forno}}, \bibinfo {author}
  {\bibfnamefont {M.~B.}\ \bibnamefont {Danailov}}, \bibinfo {author}
  {\bibfnamefont {G.}~\bibnamefont {D{\textquotesingle}Auria}}, \bibinfo
  {author} {\bibfnamefont {A.}~\bibnamefont {Demidovich}}, \bibinfo {author}
  {\bibfnamefont {G.}~\bibnamefont {De~Ninno}}, \bibinfo {author}
  {\bibfnamefont {S.}~\bibnamefont {Di~Mitri}}, \emph {et~al.},\ }\bibfield
  {title} {\bibinfo {title} {Highly coherent and stable pulses from the {FERMI}
  seeded free-electron laser in the extreme ultraviolet},\ }\href
  {https://doi.org/10.1038/nphoton.2012.233} {\bibfield  {journal} {\bibinfo
  {journal} {Nat. Photonics}\ }\textbf {\bibinfo {volume} {6}},\ \bibinfo
  {pages} {699} (\bibinfo {year} {2012})}\BibitemShut {NoStop}%
\bibitem [{\citenamefont {Prat}\ \emph {et~al.}(2024)\citenamefont {Prat},
  \citenamefont {Kittel}, \citenamefont {Calvi}, \citenamefont {Craievich},
  \citenamefont {Dijkstal}, \citenamefont {Reiche}, \citenamefont
  {Schietinger},\ and\ \citenamefont {Wang}}]{Prat2024}%
  \BibitemOpen
  \bibfield  {author} {\bibinfo {author} {\bibfnamefont {E.}~\bibnamefont
  {Prat}}, \bibinfo {author} {\bibfnamefont {C.}~\bibnamefont {Kittel}},
  \bibinfo {author} {\bibfnamefont {M.}~\bibnamefont {Calvi}}, \bibinfo
  {author} {\bibfnamefont {P.}~\bibnamefont {Craievich}}, \bibinfo {author}
  {\bibfnamefont {P.}~\bibnamefont {Dijkstal}}, \bibinfo {author}
  {\bibfnamefont {S.}~\bibnamefont {Reiche}}, \bibinfo {author} {\bibfnamefont
  {T.}~\bibnamefont {Schietinger}},\ and\ \bibinfo {author} {\bibfnamefont
  {G.}~\bibnamefont {Wang}},\ }\bibfield  {title} {\bibinfo {title}
  {Experimental characterization of the optical klystron effect to measure the
  intrinsic energy spread of high-brightness electron beams},\ }\href
  {https://doi.org/10.1103/PhysRevAccelBeams.27.030701} {\bibfield  {journal}
  {\bibinfo  {journal} {Phys. Rev. Accel. Beams}\ }\textbf {\bibinfo {volume}
  {27}},\ \bibinfo {pages} {030701} (\bibinfo {year} {2024})}\BibitemShut
  {NoStop}%
\bibitem [{\citenamefont {Dijkstal}(2022)}]{Dijkstal2022b}%
  \BibitemOpen
  \bibfield  {author} {\bibinfo {author} {\bibfnamefont {P.}~\bibnamefont
  {Dijkstal}},\ }\emph {\bibinfo {title} {Temporal {FEL} Pulse Shaping and
  Diagnostics at {SwissFEL}}},\ \href
  {https://doi.org/10.3929/ethz-b-000576424} {Ph.D. thesis},\ \bibinfo
  {school} {ETH Zurich} (\bibinfo {year} {2022})\BibitemShut {NoStop}%
\bibitem [{\citenamefont {Craievich}\ and\ \citenamefont
  {Lutman}(2017)}]{Craievich2017}%
  \BibitemOpen
  \bibfield  {author} {\bibinfo {author} {\bibfnamefont {P.}~\bibnamefont
  {Craievich}}\ and\ \bibinfo {author} {\bibfnamefont {A.~A.}\ \bibnamefont
  {Lutman}},\ }\bibfield  {title} {\bibinfo {title} {Effects of the quadrupole
  wakefields in a passive streaker},\ }\href
  {https://doi.org/10.1016/j.nima.2016.10.010} {\bibfield  {journal} {\bibinfo
  {journal} {Nucl. Instrum. Methods Phys. Res., Sect. A}\ }\textbf {\bibinfo
  {volume} {865}},\ \bibinfo {pages} {55} (\bibinfo {year} {2017})}\BibitemShut
  {NoStop}%
\bibitem [{\citenamefont {Dijkstal}\ \emph {et~al.}(2023)\citenamefont
  {Dijkstal}, \citenamefont {Ganter}, \citenamefont {Heimgartner},
  \citenamefont {Malyzhenkov}, \citenamefont {Prat}, \citenamefont {Reiche},\
  and\ \citenamefont {Craievich}}]{Dijkstal2023}%
  \BibitemOpen
  \bibfield  {author} {\bibinfo {author} {\bibfnamefont {P.}~\bibnamefont
  {Dijkstal}}, \bibinfo {author} {\bibfnamefont {R.}~\bibnamefont {Ganter}},
  \bibinfo {author} {\bibfnamefont {P.}~\bibnamefont {Heimgartner}}, \bibinfo
  {author} {\bibfnamefont {A.}~\bibnamefont {Malyzhenkov}}, \bibinfo {author}
  {\bibfnamefont {E.}~\bibnamefont {Prat}}, \bibinfo {author} {\bibfnamefont
  {S.}~\bibnamefont {Reiche}},\ and\ \bibinfo {author} {\bibfnamefont
  {P.}~\bibnamefont {Craievich}},\ }\bibfield  {title} {\bibinfo {title}
  {Corrugated wakefield structures at {SwissFEL}},\ }in\ \href
  {https://doi.org/10.18429/JACoW-IPAC2023-THPL153} {\emph {\bibinfo
  {booktitle} {Proc. IPAC 2023, Venice, Italy}}},\ \bibinfo {editor} {edited
  by\ \bibinfo {editor} {\bibfnamefont {R.}~\bibnamefont {Assmann}}, \bibinfo
  {editor} {\bibfnamefont {P.}~\bibnamefont {McIntosh}}, \bibinfo {editor}
  {\bibfnamefont {G.}~\bibnamefont {Bisoffi}}, \bibinfo {editor} {\bibfnamefont
  {A.}~\bibnamefont {Fabris}}, \bibinfo {editor} {\bibfnamefont
  {I.}~\bibnamefont {Andrian}},\ and\ \bibinfo {editor} {\bibfnamefont
  {G.}~\bibnamefont {Vinicola}}}\ (\bibinfo  {publisher} {JACoW},\ \bibinfo
  {address} {Geneva},\ \bibinfo {year} {2023})\ \bibinfo {note}
  {{THPL153}}\BibitemShut {NoStop}%
\bibitem [{\citenamefont {Han}\ \emph {et~al.}(2015)\citenamefont {Han},
  \citenamefont {Hong}, \citenamefont {Lee}, \citenamefont {Chae},
  \citenamefont {Baek}, \citenamefont {Choi}, \citenamefont {Ha}, \citenamefont
  {Hu}, \citenamefont {Hwang}, \citenamefont {Jung} \emph {et~al.}}]{Han2015}%
  \BibitemOpen
  \bibfield  {author} {\bibinfo {author} {\bibfnamefont {J.}~\bibnamefont
  {Han}}, \bibinfo {author} {\bibfnamefont {J.}~\bibnamefont {Hong}}, \bibinfo
  {author} {\bibfnamefont {J.}~\bibnamefont {Lee}}, \bibinfo {author}
  {\bibfnamefont {M.}~\bibnamefont {Chae}}, \bibinfo {author} {\bibfnamefont
  {S.}~\bibnamefont {Baek}}, \bibinfo {author} {\bibfnamefont {H.}~\bibnamefont
  {Choi}}, \bibinfo {author} {\bibfnamefont {T.}~\bibnamefont {Ha}}, \bibinfo
  {author} {\bibfnamefont {J.}~\bibnamefont {Hu}}, \bibinfo {author}
  {\bibfnamefont {W.}~\bibnamefont {Hwang}}, \bibinfo {author} {\bibfnamefont
  {S.}~\bibnamefont {Jung}}, \emph {et~al.},\ }\bibfield  {title} {\bibinfo
  {title} {Beam operation of the {PAL-XFEL} injector test facility},\ }in\
  \href {https://accelconf.web.cern.ch/FEL2014/papers/web02.pdf} {\emph
  {\bibinfo {booktitle} {Proc. FEL 2014, Basel, Switzerland}}},\ \bibinfo
  {editor} {edited by\ \bibinfo {editor} {\bibfnamefont {J.}~\bibnamefont
  {Chrin}}, \bibinfo {editor} {\bibfnamefont {S.}~\bibnamefont {Reiche}},\ and\
  \bibinfo {editor} {\bibfnamefont {V.~R.~W.}\ \bibnamefont {Schaa}}}\
  (\bibinfo  {publisher} {JACoW},\ \bibinfo {address} {Geneva},\ \bibinfo
  {year} {2015})\ \bibinfo {note} {{WEB02}}\BibitemShut {NoStop}%
\bibitem [{\citenamefont {Tomin}\ \emph {et~al.}(2023)\citenamefont {Tomin},
  \citenamefont {Schneidmiller},\ and\ \citenamefont {Decking}}]{Tomin2023}%
  \BibitemOpen
  \bibfield  {author} {\bibinfo {author} {\bibfnamefont {S.}~\bibnamefont
  {Tomin}}, \bibinfo {author} {\bibfnamefont {E.}~\bibnamefont
  {Schneidmiller}},\ and\ \bibinfo {author} {\bibfnamefont {W.}~\bibnamefont
  {Decking}},\ }\bibfield  {title} {\bibinfo {title} {First measurement of
  energy diffusion in an electron beam due to quantum fluctuations in the
  undulator radiation},\ }\bibfield  {journal} {\bibinfo  {journal} {Scientific
  Reports}\ }\textbf {\bibinfo {volume} {13}},\ \href
  {https://doi.org/10.1038/s41598-023-28813-8} {10.1038/s41598-023-28813-8}
  (\bibinfo {year} {2023})\BibitemShut {NoStop}%
\bibitem [{\citenamefont {Bane}\ \emph
  {et~al.}(2016{\natexlab{a}})\citenamefont {Bane}, \citenamefont {Stupakov},\
  and\ \citenamefont {Zagorodnov}}]{Bane2016}%
  \BibitemOpen
  \bibfield  {author} {\bibinfo {author} {\bibfnamefont {K.}~\bibnamefont
  {Bane}}, \bibinfo {author} {\bibfnamefont {G.}~\bibnamefont {Stupakov}},\
  and\ \bibinfo {author} {\bibfnamefont {I.}~\bibnamefont {Zagorodnov}},\
  }\bibfield  {title} {\bibinfo {title} {Analytical formulas for short bunch
  wakes in a flat dechirper},\ }\href
  {https://doi.org/10.1103/physrevaccelbeams.19.084401} {\bibfield  {journal}
  {\bibinfo  {journal} {Phys. Rev. Accel. Beams}\ }\textbf {\bibinfo {volume}
  {19}},\ \bibinfo {pages} {084401} (\bibinfo {year}
  {2016}{\natexlab{a}})}\BibitemShut {NoStop}%
\bibitem [{\citenamefont {Bane}\ \emph
  {et~al.}(2016{\natexlab{b}})\citenamefont {Bane}, \citenamefont {Stupakov},\
  and\ \citenamefont {Zagorodnov}}]{Bane2016a}%
  \BibitemOpen
  \bibfield  {author} {\bibinfo {author} {\bibfnamefont {K.}~\bibnamefont
  {Bane}}, \bibinfo {author} {\bibfnamefont {G.}~\bibnamefont {Stupakov}},\
  and\ \bibinfo {author} {\bibfnamefont {I.}~\bibnamefont {Zagorodnov}},\
  }\href {https://www.slac.stanford.edu/pubs/slacpubs/16750/slac-pub-16881.pdf}
  {\emph {\bibinfo {title} {{W}akefields of a {B}eam near a {S}ingle {P}late in
  a {F}lat {D}echirper}}},\ \bibinfo {type} {SLAC-PUB}\ \bibinfo {number}
  {16881}\ (\bibinfo  {institution} {Stanford Linear Accelerator Center},\
  \bibinfo {year} {2016})\BibitemShut {NoStop}%
\bibitem [{\citenamefont {Lockmann}\ \emph {et~al.}(2020)\citenamefont
  {Lockmann}, \citenamefont {Gerth}, \citenamefont {Schmidt},\ and\
  \citenamefont {Wesch}}]{Lockmann2020}%
  \BibitemOpen
  \bibfield  {author} {\bibinfo {author} {\bibfnamefont {N.~M.}\ \bibnamefont
  {Lockmann}}, \bibinfo {author} {\bibfnamefont {C.}~\bibnamefont {Gerth}},
  \bibinfo {author} {\bibfnamefont {B.}~\bibnamefont {Schmidt}},\ and\ \bibinfo
  {author} {\bibfnamefont {S.}~\bibnamefont {Wesch}},\ }\bibfield  {title}
  {\bibinfo {title} {Noninvasive thz spectroscopy for bunch current profile
  reconstructions at {MHz} repetition rates},\ }\href
  {https://doi.org/10.1103/physrevaccelbeams.23.112801} {\bibfield  {journal}
  {\bibinfo  {journal} {Phys. Rev. Accel. Beams}\ }\textbf {\bibinfo {volume}
  {23}},\ \bibinfo {pages} {112801} (\bibinfo {year} {2020})}\BibitemShut
  {NoStop}%
\bibitem [{\citenamefont {Meykopff}\ and\ \citenamefont
  {Beutner}(2018)}]{Meykopff2018}%
  \BibitemOpen
  \bibfield  {author} {\bibinfo {author} {\bibfnamefont {S.}~\bibnamefont
  {Meykopff}}\ and\ \bibinfo {author} {\bibfnamefont {B.}~\bibnamefont
  {Beutner}},\ }\bibfield  {title} {\bibinfo {title} {Emittance measurement and
  optics matching at the {European XFEL}},\ }in\ \href
  {https://doi.org/10.18429/JACOW-ICALEPCS2017-THPHA116} {\emph {\bibinfo
  {booktitle} {Proc. ICALEPCS 2017, Barcelona, Spain}}},\ \bibinfo {editor}
  {edited by\ \bibinfo {editor} {\bibfnamefont {I.}~\bibnamefont {Costa}},
  \bibinfo {editor} {\bibfnamefont {D.}~\bibnamefont {Fern{\'{a}}ndez}},
  \bibinfo {editor} {\bibfnamefont {{\'{O}}.}~\bibnamefont {Matilla}},\ and\
  \bibinfo {editor} {\bibfnamefont {V.~R.~W.}\ \bibnamefont {Schaa}}}\
  (\bibinfo  {publisher} {JACoW},\ \bibinfo {address} {Geneva},\ \bibinfo
  {year} {2018})\ \bibinfo {note} {{THPHA116}}\BibitemShut {NoStop}%
\bibitem [{\citenamefont {Maltezopoulos}\ \emph {et~al.}(2019)\citenamefont
  {Maltezopoulos}, \citenamefont {Dietrich}, \citenamefont {Freund},
  \citenamefont {Jastrow}, \citenamefont {Koch}, \citenamefont {Laksman},
  \citenamefont {Liu}, \citenamefont {Planas}, \citenamefont {Sorokin},
  \citenamefont {Tiedtke} \emph {et~al.}}]{Maltezopoulos2019}%
  \BibitemOpen
  \bibfield  {author} {\bibinfo {author} {\bibfnamefont {T.}~\bibnamefont
  {Maltezopoulos}}, \bibinfo {author} {\bibfnamefont {F.}~\bibnamefont
  {Dietrich}}, \bibinfo {author} {\bibfnamefont {W.}~\bibnamefont {Freund}},
  \bibinfo {author} {\bibfnamefont {U.~F.}\ \bibnamefont {Jastrow}}, \bibinfo
  {author} {\bibfnamefont {A.}~\bibnamefont {Koch}}, \bibinfo {author}
  {\bibfnamefont {J.}~\bibnamefont {Laksman}}, \bibinfo {author} {\bibfnamefont
  {J.}~\bibnamefont {Liu}}, \bibinfo {author} {\bibfnamefont {M.}~\bibnamefont
  {Planas}}, \bibinfo {author} {\bibfnamefont {A.~A.}\ \bibnamefont {Sorokin}},
  \bibinfo {author} {\bibfnamefont {K.}~\bibnamefont {Tiedtke}}, \emph
  {et~al.},\ }\bibfield  {title} {\bibinfo {title} {Operation of {X}-ray gas
  monitors at the {European XFEL}},\ }\href
  {https://doi.org/10.1107/S1600577519003795} {\bibfield  {journal} {\bibinfo
  {journal} {J. Synchrotron Radiat.}\ }\textbf {\bibinfo {volume} {26}},\
  \bibinfo {pages} {1045} (\bibinfo {year} {2019})}\BibitemShut {NoStop}%
\bibitem [{\citenamefont {Qin}\ \emph {et~al.}(2017)\citenamefont {Qin},
  \citenamefont {Ding}, \citenamefont {Lutman},\ and\ \citenamefont
  {Chao}}]{Qin2017}%
  \BibitemOpen
  \bibfield  {author} {\bibinfo {author} {\bibfnamefont {W.}~\bibnamefont
  {Qin}}, \bibinfo {author} {\bibfnamefont {Y.}~\bibnamefont {Ding}}, \bibinfo
  {author} {\bibfnamefont {A.~A.}\ \bibnamefont {Lutman}},\ and\ \bibinfo
  {author} {\bibfnamefont {Y.-C.}\ \bibnamefont {Chao}},\ }\bibfield  {title}
  {\bibinfo {title} {Matching-based fresh-slice method for generating two-color
  {X}-ray free-electron lasers},\ }\href
  {https://doi.org/10.1103/physrevaccelbeams.20.090701} {\bibfield  {journal}
  {\bibinfo  {journal} {Phys. Rev. Accel. Beams}\ }\textbf {\bibinfo {volume}
  {20}},\ \bibinfo {pages} {090701} (\bibinfo {year} {2017})}\BibitemShut
  {NoStop}%
\bibitem [{\citenamefont {Ratner}\ \emph {et~al.}(2015)\citenamefont {Ratner},
  \citenamefont {Behrens}, \citenamefont {Ding}, \citenamefont {Huang},
  \citenamefont {Marinelli}, \citenamefont {Maxwell},\ and\ \citenamefont
  {Zhou}}]{Ratner2015}%
  \BibitemOpen
  \bibfield  {author} {\bibinfo {author} {\bibfnamefont {D.}~\bibnamefont
  {Ratner}}, \bibinfo {author} {\bibfnamefont {C.}~\bibnamefont {Behrens}},
  \bibinfo {author} {\bibfnamefont {Y.}~\bibnamefont {Ding}}, \bibinfo {author}
  {\bibfnamefont {Z.}~\bibnamefont {Huang}}, \bibinfo {author} {\bibfnamefont
  {A.}~\bibnamefont {Marinelli}}, \bibinfo {author} {\bibfnamefont
  {T.}~\bibnamefont {Maxwell}},\ and\ \bibinfo {author} {\bibfnamefont
  {F.}~\bibnamefont {Zhou}},\ }\bibfield  {title} {\bibinfo {title}
  {Time-resolved imaging of the microbunching instability and energy spread at
  the {L}inac {C}oherent {L}ight {S}ource},\ }\href
  {https://doi.org/10.1103/PhysRevSTAB.18.030704} {\bibfield  {journal}
  {\bibinfo  {journal} {Phys. Rev. ST Accel. Beams}\ }\textbf {\bibinfo
  {volume} {18}},\ \bibinfo {pages} {030704} (\bibinfo {year}
  {2015})}\BibitemShut {NoStop}%
\bibitem [{\citenamefont {Prat}\ \emph
  {et~al.}(2020{\natexlab{b}})\citenamefont {Prat}, \citenamefont {Dijkstal},
  \citenamefont {Ferrari}, \citenamefont {Malyzhenkov},\ and\ \citenamefont
  {Reiche}}]{Prat2020c}%
  \BibitemOpen
  \bibfield  {author} {\bibinfo {author} {\bibfnamefont {E.}~\bibnamefont
  {Prat}}, \bibinfo {author} {\bibfnamefont {P.}~\bibnamefont {Dijkstal}},
  \bibinfo {author} {\bibfnamefont {E.}~\bibnamefont {Ferrari}}, \bibinfo
  {author} {\bibfnamefont {A.}~\bibnamefont {Malyzhenkov}},\ and\ \bibinfo
  {author} {\bibfnamefont {S.}~\bibnamefont {Reiche}},\ }\bibfield  {title}
  {\bibinfo {title} {High-resolution dispersion-based measurement of the
  electron beam energy spread},\ }\href
  {https://doi.org/10.1103/physrevaccelbeams.23.090701} {\bibfield  {journal}
  {\bibinfo  {journal} {Phys. Rev. Accel. Beams}\ }\textbf {\bibinfo {volume}
  {23}},\ \bibinfo {pages} {090701} (\bibinfo {year}
  {2020}{\natexlab{b}})}\BibitemShut {NoStop}%
\bibitem [{\citenamefont {Agapov}\ \emph {et~al.}(2014)\citenamefont {Agapov},
  \citenamefont {Geloni}, \citenamefont {Tomin},\ and\ \citenamefont
  {Zagorodnov}}]{Agapov2014}%
  \BibitemOpen
  \bibfield  {author} {\bibinfo {author} {\bibfnamefont {I.}~\bibnamefont
  {Agapov}}, \bibinfo {author} {\bibfnamefont {G.}~\bibnamefont {Geloni}},
  \bibinfo {author} {\bibfnamefont {S.}~\bibnamefont {Tomin}},\ and\ \bibinfo
  {author} {\bibfnamefont {I.}~\bibnamefont {Zagorodnov}},\ }\bibfield  {title}
  {\bibinfo {title} {{OCELOT}: A software framework for synchrotron light
  source and {FEL} studies},\ }\href
  {https://doi.org/10.1016/j.nima.2014.09.057} {\bibfield  {journal} {\bibinfo
  {journal} {Nucl. Instrum. Methods Phys. Res., Sect. A}\ }\textbf {\bibinfo
  {volume} {768}},\ \bibinfo {pages} {151} (\bibinfo {year}
  {2014})}\BibitemShut {NoStop}%
\bibitem [{\citenamefont {Dijkstal}(2024)}]{Dijkstal2024}%
  \BibitemOpen
  \bibfield  {author} {\bibinfo {author} {\bibfnamefont {P.}~\bibnamefont
  {Dijkstal}},\ }\bibfield  {title} {\bibinfo {title} {Longitudinal phase space
  diagnostics with a non-movable corrugated passive wakefield streaker},\
  }\href {https://doi.org/10.5281/zenodo.10673911} {10.5281/zenodo.10673911}
  (\bibinfo {year} {2024})\BibitemShut {NoStop}%
\bibitem [{\citenamefont {Bane}\ and\ \citenamefont
  {Stupakov}(2016)}]{Bane2016b}%
  \BibitemOpen
  \bibfield  {author} {\bibinfo {author} {\bibfnamefont {K.}~\bibnamefont
  {Bane}}\ and\ \bibinfo {author} {\bibfnamefont {G.}~\bibnamefont
  {Stupakov}},\ }\bibfield  {title} {\bibinfo {title} {Dechirper wakefields for
  short bunches},\ }\href {https://doi.org/10.1016/j.nima.2016.02.055}
  {\bibfield  {journal} {\bibinfo  {journal} {Nucl. Instrum. Methods Phys.
  Res., Sect. A}\ }\textbf {\bibinfo {volume} {820}},\ \bibinfo {pages} {156}
  (\bibinfo {year} {2016})}\BibitemShut {NoStop}%
\bibitem [{\citenamefont {Dohlus}\ and\ \citenamefont
  {Wanzenberg}(2017)}]{Dohlus2017}%
  \BibitemOpen
  \bibfield  {author} {\bibinfo {author} {\bibfnamefont {M.}~\bibnamefont
  {Dohlus}}\ and\ \bibinfo {author} {\bibfnamefont {R.}~\bibnamefont
  {Wanzenberg}},\ }\bibfield  {title} {\bibinfo {title} {An introduction to
  wake fields and impedances},\ }in\ \href
  {https://doi.org/10.23730/CYRSP-2017-003.15} {\emph {\bibinfo {booktitle}
  {Proceedings of the CAS-CERN Accelerator School on Intensity Limitations in
  Particle Beams}}},\ Vol.~\bibinfo {volume} {3},\ \bibinfo {editor} {edited
  by\ \bibinfo {editor} {\bibfnamefont {W.}~\bibnamefont {Herr}}}\ (\bibinfo
  {publisher} {CERN},\ \bibinfo {year} {2017})\BibitemShut {NoStop}%
\bibitem [{\citenamefont {Zagorodnov}\ \emph {et~al.}(2016)\citenamefont
  {Zagorodnov}, \citenamefont {Feng},\ and\ \citenamefont
  {Limberg}}]{Zagorodnov2016}%
  \BibitemOpen
  \bibfield  {author} {\bibinfo {author} {\bibfnamefont {I.}~\bibnamefont
  {Zagorodnov}}, \bibinfo {author} {\bibfnamefont {G.}~\bibnamefont {Feng}},\
  and\ \bibinfo {author} {\bibfnamefont {T.}~\bibnamefont {Limberg}},\
  }\bibfield  {title} {\bibinfo {title} {Corrugated structure insertion for
  extending the {SASE} bandwidth up to 3\% at the {European XFEL}},\ }\href
  {https://doi.org/10.1016/j.nima.2016.09.001} {\bibfield  {journal} {\bibinfo
  {journal} {Nucl. Instrum. Methods Phys. Res., Sect. A}\ }\textbf {\bibinfo
  {volume} {837}},\ \bibinfo {pages} {69} (\bibinfo {year} {2016})}\BibitemShut
  {NoStop}%
\bibitem [{\citenamefont {Qin}\ \emph {et~al.}(2023)\citenamefont {Qin},
  \citenamefont {Dohlus},\ and\ \citenamefont {Zagorodnov}}]{Qin2023}%
  \BibitemOpen
  \bibfield  {author} {\bibinfo {author} {\bibfnamefont {W.}~\bibnamefont
  {Qin}}, \bibinfo {author} {\bibfnamefont {M.}~\bibnamefont {Dohlus}},\ and\
  \bibinfo {author} {\bibfnamefont {I.}~\bibnamefont {Zagorodnov}},\ }\bibfield
   {title} {\bibinfo {title} {Short-range wakefields in an {L}-shaped
  corrugated structure},\ }\href
  {https://doi.org/10.1103/physrevaccelbeams.26.064402} {\bibfield  {journal}
  {\bibinfo  {journal} {Phys. Rev. Accel. Beams}\ }\textbf {\bibinfo {volume}
  {26}},\ \bibinfo {pages} {064402} (\bibinfo {year} {2023})}\BibitemShut
  {NoStop}%
\end{thebibliography}%
\end{document}